\documentclass[a4paper,10pt]{article}
\usepackage{graphicx,hyperref}
\usepackage{amsmath,amsfonts,amssymb}
\usepackage{wasysym}
\usepackage{fullpage}
\usepackage{enumerate}
\usepackage{sidecap}
\usepackage{lscape}
\usepackage{authblk}
\usepackage{appendix}
\usepackage{setspace}
\usepackage{subfigure}
\title{\bf{Complex contagion process in spreading of online innovation}}
\author[1,2,3,4]{M\'arton Karsai \thanks{marton.karsai@ens-lyon.fr}}
\author[2]{Gerardo I\~{n}iguez}
\author[2,5]{Kimmo Kaski}
\author[6,7,2]{J\'anos Kert\'esz}
\affil[1]{\small{Laboratory for the Modeling of Biological and Socio-technical Systems, Northeastern University, Boston MA 02115 USA}}
\affil[2]{Department of Biomedical Engineering and Computational Science (BECS), Aalto University School of Science, FI-00076 AALTO, Finland}
\affil[3]{Software Technology and Applications Competence Center (STACC), 51003 Tartu, Estonia}
\affil[4]{Laboratoire de l'Informatique du Parall\'elisme, INRIA-UMR 5668, IXXI,  ENS de Lyon, 69364 Lyon, France}
\affil[5]{CABDyN Complexity Centre, S\"aid Business School, University of Oxford, OX1 1HP, UK}
\affil[6]{Center for Network Science, Central European University, H-1051 Budapest, Hungary}
\affil[7]{Institute of Physics, Budapest University of Technology and Economics, H-1111 Budapest, Hungary}

\begin{document}

\maketitle

\begin{abstract}
Diffusion of innovation can be interpreted as a social spreading phenomena governed by the impact of media and social interactions. Although these mechanisms have been identified by quantitative theories, their role and relative importance are not entirely understood, since empirical verification has so far been hindered by the lack of appropriate data. Here we analyse a dataset recording the spreading dynamics of the world's largest Voice over Internet Protocol service to empirically support the assumptions behind models of social contagion. We show that the rate of spontaneous service adoption is constant, the probability of adoption via social influence is linearly proportional to the fraction of adopting neighbours, and the rate of service termination is time-invariant and independent of the behaviour of peers. By implementing the detected diffusion mechanisms into a dynamical agent-based model, we are able to emulate the adoption dynamics of the service in several countries worldwide. This approach enables us to make medium-term predictions of service adoption and disclose dependencies between the dynamics of innovation spreading and the socioeconomic development of a country.
\end{abstract}
\vspace{.1in}
\textbf{keywords:} complex contagion phenomena, mean-field approximation, data-driven modelling
\vspace{.2in}

Diffusion of news, ideas, and innovations as well as the distribution of services and products are all examples of social spreading phenomena that have become an integral part of our everyday life, strongly accelerated by novel, web-based interaction channels. These innovations serve as the engine of economic development \cite{Solow}, but only their diffusion throughout society brings them to success. The processes involved in innovation spreading have been in the focus of research for decades \cite{Solow,Bass1,Toole2012,Onnela2010}, yet their dynamics and modelling have remained as challenges to our scientific understanding.

The propagation of innovations takes place in a social network \cite{Bass1,Toole2012,Onnela2010,Aral2009} and is driven by the entanglement of individuals' decision-making processes \cite{Ullah} as well as by the influence of media and social interactions \cite{Rogers,Wejnert}. Although the effects of network structure on contagion processes have recently been shown to be important \cite{Barrat2008}, knowledge about the social network itself is rather limited since its structure and dynamics usually remain hidden. In this respect the digital age has opened up unprecedented opportunities, as online social networks and Voice over Internet Protocol services record detailed information of the connections and activities of their users. These services partially decode the underlying social structure by acting as proxies for the network of real social ties between individuals, and also provide accurate records of the users' adoption behaviour. In this way the different sources of influence on the decisions of an individual immersed in a perpetually changing environment of social interactions become traceable. We are therefore encouraged to devise dynamic agent-based models to describe, simulate, and even predict emergent behaviour of such social contagion phenomena \cite{Vespignani2009,Jackson2005,Pintado2008}.

\begin{figure*}
\centerline{\includegraphics[width=.8\textwidth]{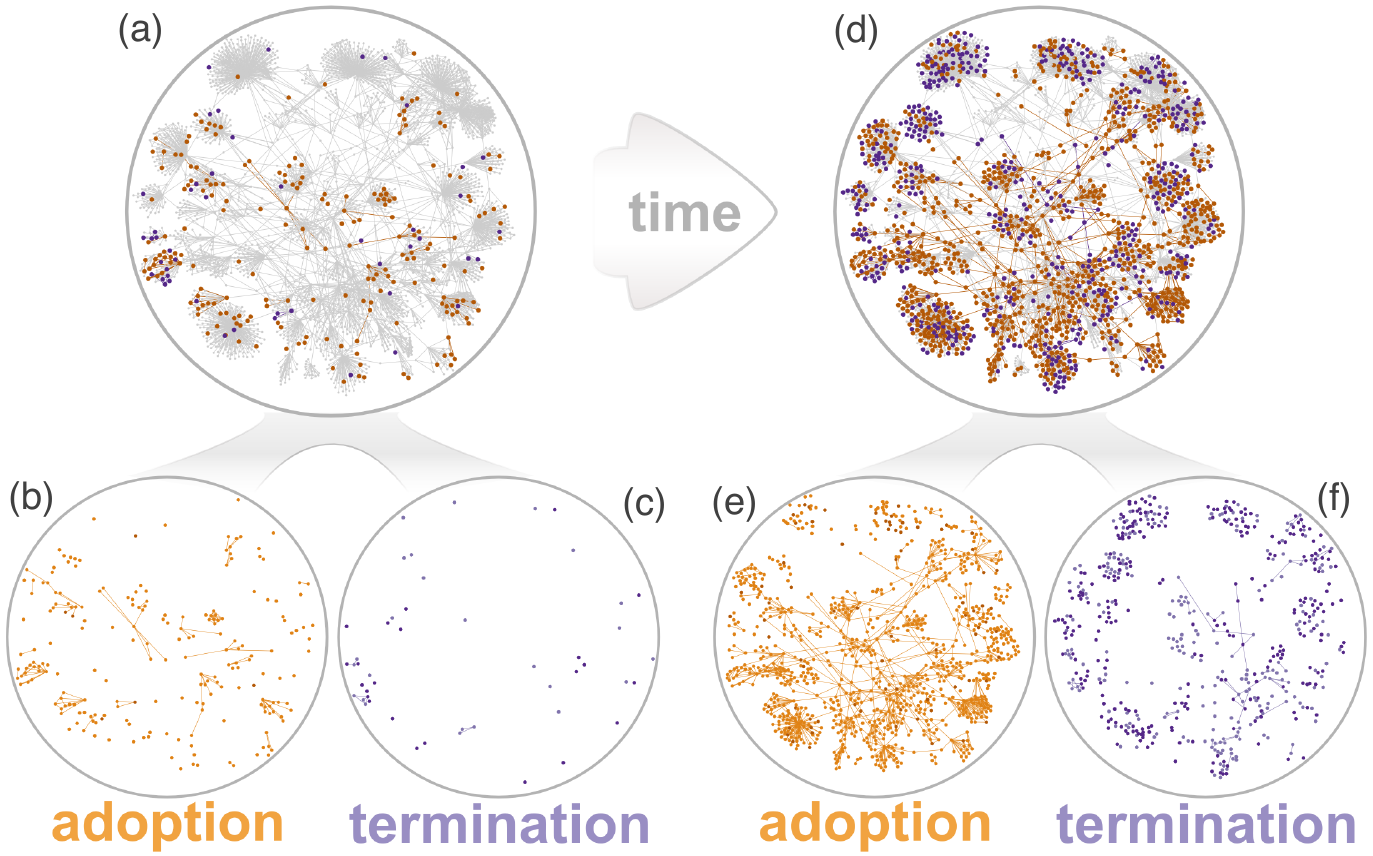}}
\caption{(Online version in colour.) Sample of the aggregated Skype network of Switzerland. (a), (d) Snowball sample (maximum distance from seed $d=4$) where user accounts and confirmed links between them are shown for two intermediate times in the adoption process. Nodes are colored according to their adoption state: grey for future users, orange for current adopters, and purple for terminated accounts (corresponding to states $S$, $A$ and $R$ in our compartmental model). A link has one of these colors if the states of the connected nodes are equal, and is grey otherwise. (b), (c), (e), (f) Decomposed networks for adoption [(b), (e)] and termination [(c), (f)]. Nodes have the same colors as before but are shaded depending on their action (dark for spontaneous action, light for peer-pressure action). Only links connecting nodes with equal states are shown. The termination network has a larger fraction of dark nodes than its adoption counterpart, meaning that social interactions affect adoption more than termination.}\label{fig:nets}
\end{figure*}

These phenomena are identified as {\it complex contagion processes} when the exposure of an individual is conditional to the decision of a fraction of its peers \cite{Centola2007}. This is particularly different from \textit{simple spreading processes}, where a rate determines the transmission of infection between nodes and one infected neighbour is always sufficient to expose a susceptible node \cite{Barrat2008,Hill}. Complex contagion phenomena are commonly modelled by processes where the fractions of adopting neighbours necessary for exposure are set as individual thresholds. This idea was first introduced by Granovetter \cite{Granovetter1978} who discussed the ideal network structure and threshold distribution to allow for the evolution of riots or other collective movements. Subsequently Watts \cite{Watts2002} proposed a simplistic model to explore sufficient structural and threshold conditions for the evolution of global adoption cascades. During the last ten years several studies contributed to the foundations of complex contagion  \cite{Watts2002,Backstrom2006,Gleeson2007,Romero2011,Bakshy2012,Singh2013}, and in addition online experiments were carried out to provide empirical evidence about the effect of social influence \cite{Centola2010,Centola2011}. Beyond the conventional threshold mechanism, the effect of homophily \cite{Aral2009,Bakshy2012,Suri2011} and the role of external media influence \cite{Toole2012} were also investigated recently.

Here we study one of today's largest online communication services, the Skype network, with over 300 million monthly connected users. Data covers the history of individuals that have adopted Skype from September 2003 until March 2011 (i.e. 2738 days), including registration events and contact network evolution for every registered user around the world. For our investigation we select user accounts with an identified country of registration and consider only their mutually confirmed connections, both within the country and abroad. To receive the best estimation of node degrees in the underlying social network, we integrate the evolving Skype network for the whole available period and count the number of confirmed relationships per node (including international ties). The adoption dynamics of a given country can be directly observed by assigning times of adoption ($t_a$) and termination ($t_t$) to all the accounts. These are respectively defined as the dates of registration and last activity (as regards to any of the services) in Skype. Explicitly, we identify any account as terminated if its last activity happened earlier than one year prior to the end of the observation period. In this way we are able to build a complete adoption and termination history of Skype for 2373 days. As an illustration of the adoption process, in Fig. \ref{fig:nets}.a and d we show a sample of the contact network of Switzerland for two intermediate times (for further details of the dataset see Supplementary Section S1).

Taking advantage of this large digital dataset, our goal is to fill the persisting gap between real observations and the assumptions made in models of product adoption spreading in techno-social networks. We empirically study the assumptions borrowed from conventional models of complex contagion and  analyse the crucial effect of social influence. Finally, we introduce an agent-based model that combines the detected diffusion mechanisms and provides plausible medium-term predictions for the spreading of online innovations in several countries worldwide.

\section*{Results}

\subsection*{The adoption dynamics}
\begin{figure}
\centerline{\includegraphics[width=.8\textwidth]{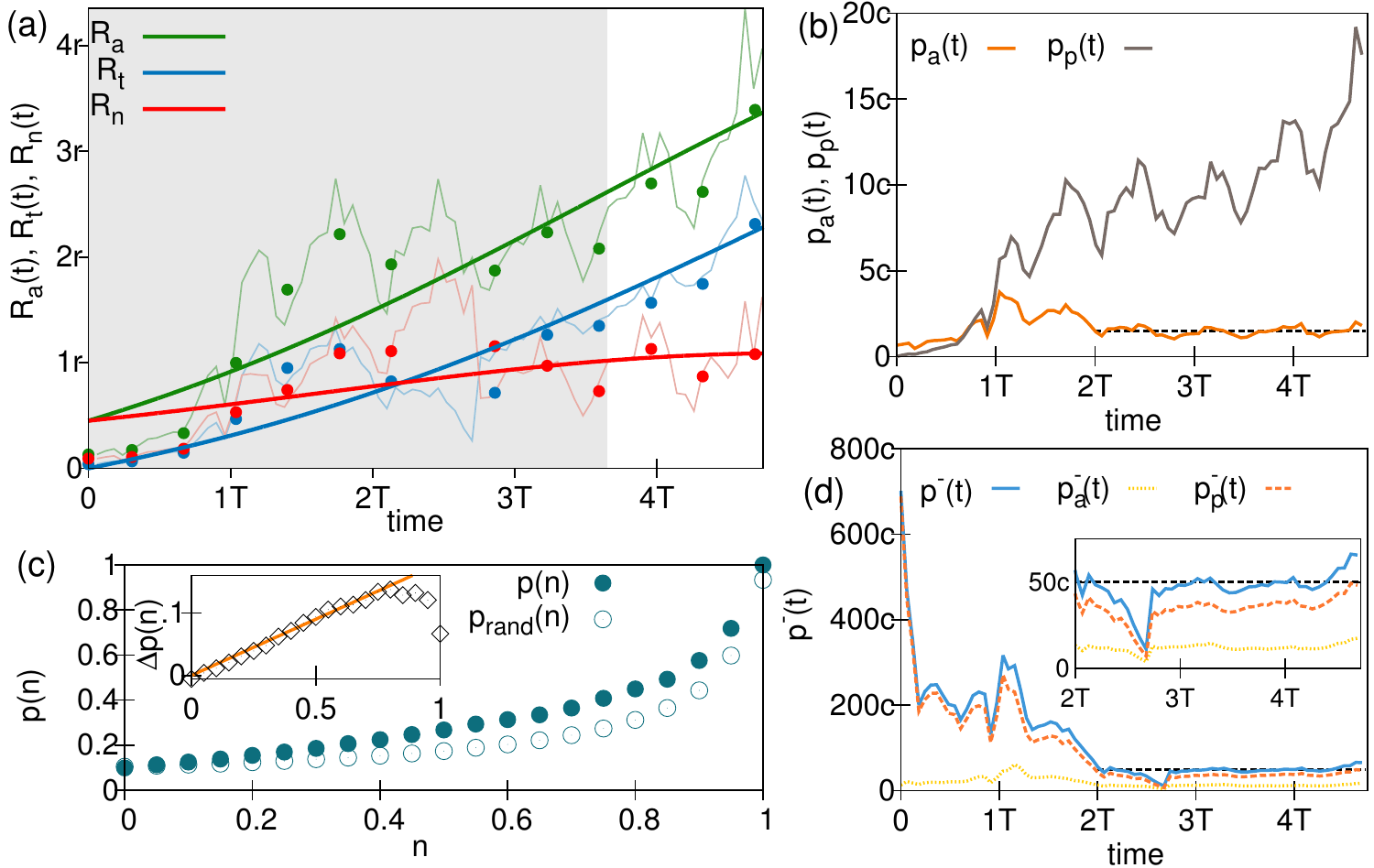}}
\caption{(Online version in colour.) Empirical rates and probabilities for Switzerland. (a) Thin curves denote empirical rates of adoption [$R_a(t)$], termination [$R_t(t)$], and net adoption [$R_n(t)$], while symbols are their corresponding binned values. A binned data point in $[2T,3T]$ has been removed due to systematic bias in $R_t(t)$ caused by a major software update during this period. A shaded (white) area indicates the training (predicted) period for the theoretical fit of our model, drawn as thick lines with the same colors as the empirical rates. (b) Probabilities of spontaneous [$p_a(t)$] and peer-pressure [$p_p(t)$] adoption per unit time. (c) Average conditional probability of adoption as a function of the fraction of adopting neighbors $n$, measured in the original data [$p(n)$, solid circles] and in the shuffled data corresponding to the null model [$p_{\mathrm{rand}}(n)$, open circles]. Inset shows the unbiased difference $\Delta p(n) = p(n) - p_{\mathrm{rand}}(n)$ (symbols) and a fitted linear function (continuous line). (d) Probabilities of overall termination [$p^-(t)$], and of spontaneous [$p_a^-(t)$] and peer-pressure [$p_p^-(t)$] termination per unit time. The inset depicts a zoom from time $2T$ onwards. $T$, $r$ and $c$ are arbitrary linear scaling constants, with time dimensions for $T$. Black lines in panels (b), (d) are fitted constants.}\label{fig:rates}
\end{figure}

The spreading of the online service is determined by competing processes of adoption and termination as described by the evolution of the corresponding rates $R_a(t)$ and $R_t(t)$, which measure the fraction of all users that adopt or terminate the service in a given time window $\Delta t$ (Fig. \ref{fig:rates}.a). These simple rate functions already disclose interesting features of the adoption dynamics, since their overall growth signals continuously accelerating processes of adoption and termination. Yet the actual time evolution of spreading service is better characterized by the net adoption rate $R_n(t)=R_a(t)-R_t(t)$ (for an overview of all empirical quantities see Table \ref{table:pars}).

Opening a user account constitutes a single event in the decision-making process of an individual that is triggered by either spontaneous decisions, the influence of media or by the social environment \cite{Aral2009,Watts2002}. On the other hand, users may terminate their accounts for several reasons including vanishing demand or dissatisfaction, by switching to another product permanently, or by simply abandoning the service with a chance of re-adoption (e.g. due to loss of password or intention for lower monitoring). Some of these processes are observable by investigating the data. An example is shown in Fig. \ref{fig:nets}, where the contact network of Switzerland is further decomposed into sub-networks of adopted and terminated users. In the former some nodes appear disconnected, which indicates individuals that have adopted Skype prior to their friends. This so-called \textit{spontaneous} adoption, where individual factors and external media play a role, is a typical adoption pattern in the beginning of the process (Fig. \ref{fig:nets}.b). Alternatively, at the time of adoption many nodes have neighbours who are already existing users, a common pattern for later times (Fig. \ref{fig:nets}.e). This second scenario of \textit{peer-pressure} adoption indicates the possible influence of the social environment. In contrast, the termination network consists mostly of single nodes at all times (Fig. \ref{fig:nets}.c and f), meaning that these users, although they are surrounded by adopters, decide individually to terminate. This observation suggests a negligible effect of social influence on termination.

\subsection*{Mechanisms of adoption}
An analysis of the evolving network structure around a given user can help us to detect whether an ego adopted or terminated the product before any of its neighbours did; or else followed the decisions previously made by a fraction of them. In this way we can label the performed action as either spontaneous or driven by peer pressure. To define the related measures we consider the underlying social network as static, meaning that its evolution requires a much larger temporal scale than the adoption process itself. This static structure is defined as the aggregated social network of Skype at the end of the recorded period, and provides a lower estimate for the total number of friends of each individual. Moreover, we assume that the maximum size of the static social network is the number $I$ of internet users in a given country at the end of the observation period (2011) \cite{Internet}, and thus define $I - N_a(t)$ as the population that has not yet adopted Skype at time $t$.

Under these assumptions, the probabilities per unit time that a user adopts either spontaneously or due to peer pressure are defined as,
\begin{equation}
\label{eq:empAdopt}
p_a(t)=\dfrac{\# ad(t+\Delta t|SF=0)}{I-N_a(t)}, \quad \text{and} \quad
p_p(t)=\dfrac{\# ad(t+\Delta t|SF\neq 0)}{I-N_a(t)},
\end{equation}
where $\# ad(t+\Delta t|SF=0)$ [$\# ad(t+\Delta t|SF\neq 0)$] is the number of users who adopt the service in a time window $\Delta t$, under the condition that their number of adopting neighbours at time $t$ is $SF=0$ ($SF \neq 0$). In a similar fashion, the probabilities per unit time that a user terminates the service either spontaneously or due to peer pressure are,
\begin{equation}
\label{eq:empTerm}
p_a^-(t) = \dfrac{\# tr(t+\Delta t,TF=0)}{N_a(t)}, \quad \text{and} \quad
p_p^-(t) = \dfrac{\# tr(t+\Delta t,TF\neq0)}{N_a(t)},
\end{equation}
where $TF$ stands for the number of neighbours of a user that have terminated usage up to time $t$ (for a discussion on the restrictions of these empirical quantities see Supplementary Section S2.1-S2.4).

The data shows that after an initial, transient period, the rate of spontaneous adoption $p_a(t)$ (Fig. \ref{fig:rates}.b) and the rate of termination $p^-(t)=p_a^-(t)+p_p^-(t)$ (Fig. \ref{fig:rates}.d) become constant apart from small fluctuations. The same holds separately for the rates of spontaneous [$p_a^-(t)$] and peer-pressure [$p_p^-(t)$] termination. The time invariance of these rates is an obvious assumption for most biological epidemics, which, however, has never been empirically shown before in the case of social contagion phenomena, despite its wide use \cite{Vespignani_review,Castellano2009}. Our results provide the first validation of this quite fundamental assumption used in the conventional modelling of social spreading processes, where probabilities analogous to the ones described here are treated like constants at the outset.

When the ego is not the first adopter among neighbours, the rate $p_p(t)$ of adoption via peer pressure is not constant but increases with time (Fig. \ref{fig:rates}.b). This is mainly due to social influence arising from the user's social circle. An appropriate way to quantify such effects is to measure the conditional probability $p(n)$ of adoption provided that a fraction $n$ of the ego's neighbours have adopted the product before as

\begin{equation}
\label{eq:pn}
p(n)=\dfrac{\# ad(n)}{N-\sum_{m=0}^{m<n}\# ad(m)}.
\end{equation}
Here the numerator counts the number of users with a fraction $n$ of adopter friends at the time of adoption, while the denominator is the number of people with a larger or equal fraction $m \geq n$, i.e. all individuals who had the chance to adopt Skype while having a fraction $n$ of adopter neighbours (for further details see Supplementary Section S2.3). We observe that the probability $p(n)$ is monotonically increasing (Fig. \ref{fig:rates}.c), an empirical finding in agreement with the assumptions of several threshold models for epidemic spreading and social dynamics \cite{Watts2002,Dodds2004,Klimek,Takaguchi2013}. However, since we cannot see the entire social network (only the part uncovered by the Skype graph), this probability is biased as $n \rightarrow 1$. To estimate such bias, we build a reference null model by shuffling the adoption times of all accounts and measuring the corresponding conditional probability $p_{\mathrm{rand}}(n)$ for this system. The shuffling procedure removes the effect of social influence but conserves the adoption rates and keeps the social structure unchanged. In other words, the reference probability is biased in the same way as the original measurement, but is not driven by social influence as all such correlations have been removed by the shuffling. Consequently the difference $\Delta p(n)= p(n) - p_{\mathrm{rand}}(n)$ quantifies the effect of social influence in the adoption process (inset of Fig. \ref{fig:rates}.c): $\Delta p(n)$ increases approximately in a linear fashion with the fraction of adopting neighbours. This observation is in agreement with previous studies where a similar scaling of social influence has been recognized through small scale experiments \cite{Centola2010}, data-driven observations \cite{Bakshy2012}, and modelling \cite{Dodds2004,Dodds2013}.

\begin{table}
\center
\begin{tabular}{| l | p{0.71\textwidth} |}
\hline 
{\bf quantity} & {\bf description} \\ \hline \hline
\multicolumn{2}{|l|}{{\it data/model quantities}} \\ \hline
$R_a(t)$, $R_t(t)$ & {\small Fraction of all users that adopt/terminate the service in a time window} \\ \hline
$R_n(t)=R_a(t)-R_t(t)$ & {\small Net rate of adoption} \\ \hline
\multicolumn{2}{|l|}{{\it empirical quantities}} \\ \hline
$t_a$, $t_t$ & {\small Time of adoption/termination of service for a user} \\ \hline
$t_l = t_t - t_a$ & {\small Lifetime of a user account in the service} \\ \hline
$\Delta t$ & {\small Time window used to define rates} \\ \hline
$I$ & {\small Number of internet users in a country at the end of observation period} \\ \hline
$N_a(t)$ & {\small Number of service users at time $t$} \\ \hline
$p_a(t)$, $p_p(t)$ & {\small Rate of spontaneous/peer-pressure adoption at time $t$} \\ \hline
$p^-_a(t)$, $p^-_p(t)$ & {\small Rate of spontaneous/peer-pressure termination at time $t$} \\ \hline
$p^-(t) = p^-_a(t) + p^-_p(t)$ & {\small Rate of termination at time $t$} \\ \hline
$p(n)$ & {\small Probability that a user adopts, if a fraction $n$ of its neighbours are adopters} \\ \hline
$p_{\mathrm{rand}}(n)$ & {\small Value of $p(n)$ after shuffling adoption times} \\ \hline
$\Delta p(n)= p(n) - p_{\mathrm{rand}}(n)$ & {\small Unbiased measure of social influence in adoption} \\ \hline
$\tau$ & {\small Inverse speed of innovation diffusion} \\ \hline
\multicolumn{2}{|l|}{{\it model parameters}} \\ \hline
$p_a$, $p_p$ & {\small Constant probability of spontaneous/peer-pressure adoption} \\ \hline
$p_s$, $p_r$ & {\small Constant probability of temporary/permanent termination} \\ \hline
$\langle k \rangle$ & {\small Average degree of network} \\ \hline
$p_{pk} = p_p(\langle k \rangle - 1)/\langle k \rangle$ & {\small Peer-pressure adoption probability, weighted by degree} \\ \hline
\multicolumn{2}{|l|}{{\it estimated parameters}} \\ \hline
$\widetilde{\langle k \rangle}, \widetilde{p^-}$ & {\small Estimates of average degree and asymptotic rate of termination} \\ \hline
$\widetilde{p_s}=(\widetilde{p^-}-p_r)/(1-p_r)$ & {\small Estimate of temporary termination probability} \\ \hline
\end{tabular} 
\caption{Data and model quantities, including notation and a brief description of their meaning. Groups correspond to: quantities measured in both the data and model; empirical quantities; parameters used by the model; and parameters estimated from the data.}
\label{table:pars}
\end{table}

\subsection*{The model process}
The analogy between epidemic spreading and social contagion has been widely used to model various societal diffusion processes \cite{Jackson2005,Hill,GoffmanNevill,DaleyKendall}. Here we take this approach to build a compartmental model based on the identified mechanisms in Skype usage, aimed at a generic description of the large-scale adoption dynamics of technological innovations. We depict individuals as agents in one of three non-overlapping states, susceptible ($S$), adopter ($A$) and removed ($R$), describing people who may adopt the product later, are users already, and will never use it again. In accordance with our observations, the behaviour of an agent can be characterized by four elementary processes: (a) \textit{Spontaneous adoption}, influenced by individual factors or external media independently of the social network. This is certainly the dominant mechanism for agents with no user neighbours at the time of adoption. (b) \textit{Peer-pressure adoption}, an intrinsic social effect implemented here by making use of the observed linear scaling of the probability $p(n)$. (c) \textit{Temporary termination}, describing the case in which agents stop usage with a chance of re-adoption. (d) \textit{Permanent termination}, when users abandon the service altogether. The flow $S \rightarrow A$ is regulated by processes (a) and (b), $A \rightarrow S$ by (c), and $A \rightarrow R$ by (d). Finally we assume that the underlying social network evolves with a much longer time scale than the ongoing adoption process, so that its structure may be considered static with fixed size.  

For large systems, the modelled adoption process can be well characterized by a rate equation formalism using the heterogeneous mean-field approximation \cite{Barrat2008,Satorras2001} (see Appendix A). This approach takes agents with identical degree to be statistically equivalent and ignores fluctuations in their dynamical properties. Thus, assuming no degree-degree correlations in the network, the adoption dynamics is reduced to the following system of non-linear ordinary differential equations,
\begin{eqnarray}
\label{eq:uncorrODEs1}
da/dt & = & [p_a + p_{pk}(1 - p_a)a]s - [p_r + p_s(1 - p_r)]a, \\
\label{eq:uncorrODEs2}
ds/dt & = & -[p_a + p_{pk}(1 - p_a)a]s + p_s(1 - p_r)a, \\
\label{eq:uncorrODEs3}
dr/dt & = & p_ra,
\end{eqnarray}
where $s(t)$, $a(t)$, and $r(t)$ are the average probabilities that an agent is in state $S$, $A$ or $R$, respectively, and satisfy the normalization condition $s(t)+a(t)+r(t)=1$. The elementary mechanisms (a--d) are parametrized through the constant probabilities of spontaneous ($p_a$) and peer-pressure ($p_p$) adoption, and of temporary ($p_s$) and permanent ($p_r$) termination (for an overview of all model parameters see Table \ref{table:pars}). Under the above conditions the model does not depend on the degree distribution of the social network, since $p_p$ appears only in the weighted form $p_{pk} = p_p(\langle k \rangle - 1)/\langle k \rangle$, with $\langle k \rangle$ the average degree of the network. Moreover, for large $\langle k \rangle$ the model becomes independent of this quantity as $p_{pk} \sim p_p$. The system (\ref{eq:uncorrODEs1})--(\ref{eq:uncorrODEs3}) finally allows us to write the theoretical rates of adoption and termination as,
\begin{eqnarray}
\label{eq:uncorrRates1}
R_a(t) & = & [p_a+p_{pk}(1-p_a)a]s, \\
\label{eq:uncorrRates2}
R_t(t) & = & [p_r+p_s(1-p_r)]a,
\end{eqnarray}
that is, the gain and loss terms in Eq.~(\ref{eq:uncorrODEs1}) (detailed derivation in Supplementary Section S3).

In order to measure the effect of degree-degree correlations on the spreading dynamics, we perform data-driven simulations by evaluating the model process over the integrated Skype network of a country (Fig. \ref{fig:model}.a). While this empirical network retains its full topological complexity (in terms of real community structure, assortativity, etc.), we consider an model scale-free network \cite{barabasi1999} of the same size and average degree as a control case. We then run the model process over the empirical and control networks, and compare their corresponding rates of adoption and termination with the mean-field prediction of Eqs.~(\ref{eq:uncorrRates1}) and~(\ref{eq:uncorrRates2}). Since the average degree of the Skype network is not too small, deviations of the simulated rates from the theoretical values are not large, resulting in rates with the same qualitative behaviour. More interestingly, there is only a small discrepancy between the rates of the empirical and control networks. This suggests that topological correlations have a minor slowing-down effect on the spreading dynamics, and thus play a negligible role in the overall rates of the adoption process. Note that a similar independence of the population structure has been observed earlier in controlled experiments of networked public goods games \cite{Suri2011} and a spatial Prisoner's Dilemma \cite{gracia2012}. These observations validate retrospectively the theoretical considerations mentioned above, where we have assumed the background social network to be uncorrelated.

\subsection*{Validation and socio-economic correlations}
In order to validate our model process, we compare the theoretical rate functions of Eqs.~(\ref{eq:uncorrRates1}) and~(\ref{eq:uncorrRates2}) with the empirical data by estimating some of the model parameters. An estimate $\widetilde{\langle k \rangle}$ for the average degree of the social network can be obtained from the fully aggregated Skype network of a country, if we consider the ego-network of each user with international links included. The estimated rate of termination, $\widetilde{p^-}$, can be measured from the long time behaviour of the spreading process (Fig. \ref{fig:rates}.d) and then used to fix $p_s$ to the value $\widetilde{p_s}=(\widetilde{p^-}-p_r)/(1-p_r)$, where $p_r$ is a free parameter. While $p_a$ could also be measured directly (by counting adoption events where no user neighbours are present at the time), the observation time of the dataset is not sufficiently long to estimate a constant value $\widetilde{p_a}$ in all countries. Therefore we leave $p_a$, $p_p$ and $p_r$ as free quantities to be fitted (estimation of $p_a$ for selected countries in Supplementary Section S4).

\begin{figure}
\centerline{\includegraphics[width=.6\textwidth]{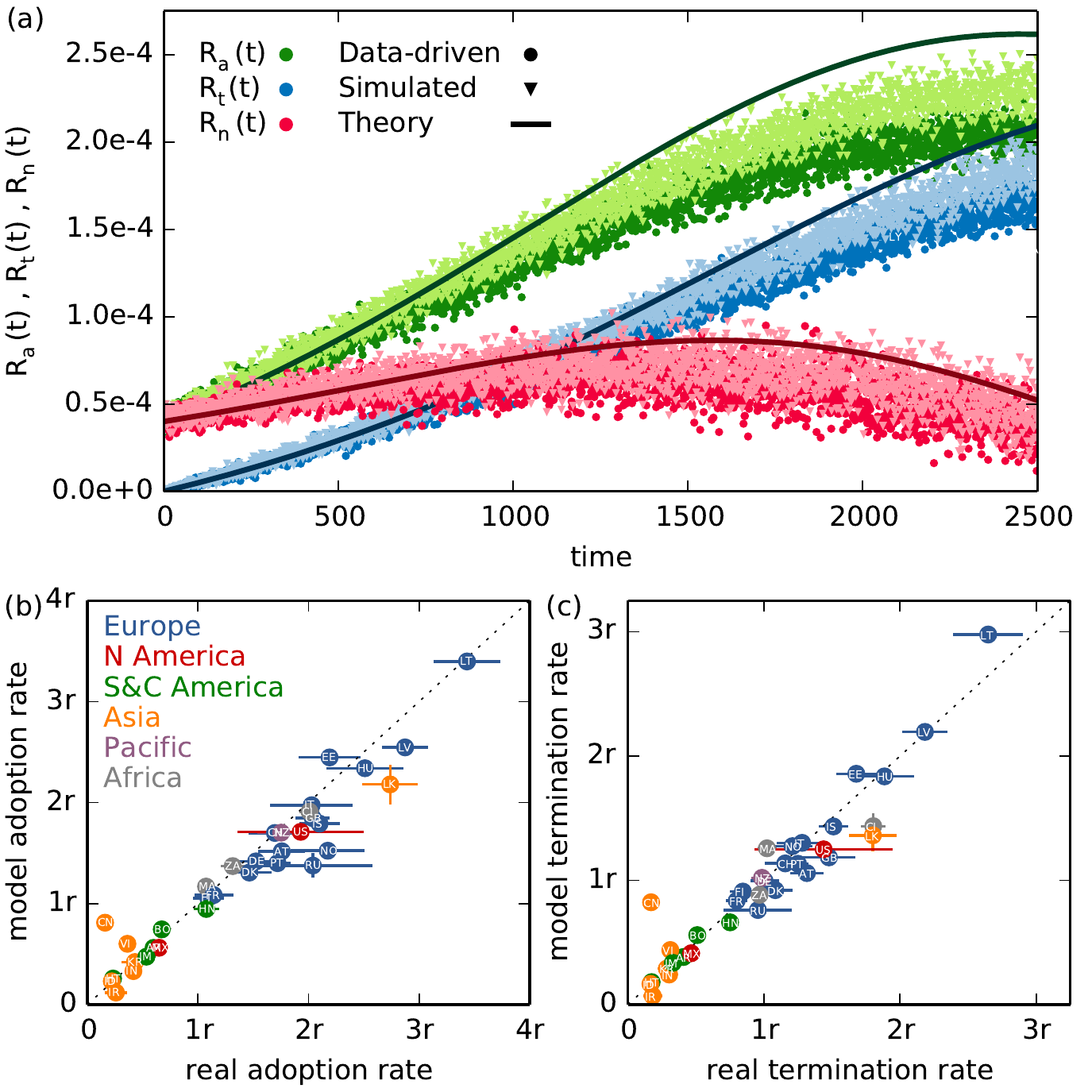}}
\caption{(Online version in colour.) Data-driven simulations and validation of model predictions. (a) Rates of the model process over the integrated Skype network of Switzerland (dark symbols) and over an uncorrelated scale-free network (light symbols) with similar size and average degree. Parameter values for the numerical simulations are arbitrarily set to $p_a = 0.00004$, $p_p = 0.0024$, $p_s = 0.0004$ and $p_r = 0.0008$. Solid lines indicate the theoretical rates calculated from Eqs.~(\ref{eq:uncorrRates1}) and~(\ref{eq:uncorrRates2}). Similar scaling of the data-driven, simulated, and theoretical curves suggests a minor effect of structural correlations. (b), (c) Comparison between empirical and theoretical values of the rates of adoption and termination for 34 different countries. Symbols depict rates averaged over the last six months of the observation period, with their corresponding standard deviations as errorbars. In-symbol letters are country abbreviations, colors denote continental territories, $r$ is an arbitrary linear scaling constant, and the dashed line is a linear function with unit slope.}\label{fig:model}
\end{figure}

Overall, the model dynamics is characterized by $\{ p_a, p_p, p_r, \widetilde{p_s}, \widetilde{\langle k \rangle} \}$, a set of three free parameters and two estimated quantities. The free parameters are used to simultaneously fit the model rates on the binned empirical rates $R_a(t)$, $R_t(t)$ and $R_n(t)$, by means of a bounded non-linear least-squares method. To ascertain the predictive power of our model, we fit over a 5-year training period and look for predictions in the last 1.5 years (Fig. \ref{fig:rates}.a). Such prognosis can be quantified by comparing the average rates provided by the model with their corresponding empirical values during the final six months of observation. After repeating the calculations for 34 different countries (with diverse levels of technological development), the related values of the final empirical and modelled rates all collapse close to a line with unit slope (Fig. \ref{fig:model}.b and c), thus validating our model for the studied adoption process.

\begin{figure}
\centerline{\includegraphics[width=.8\textwidth]{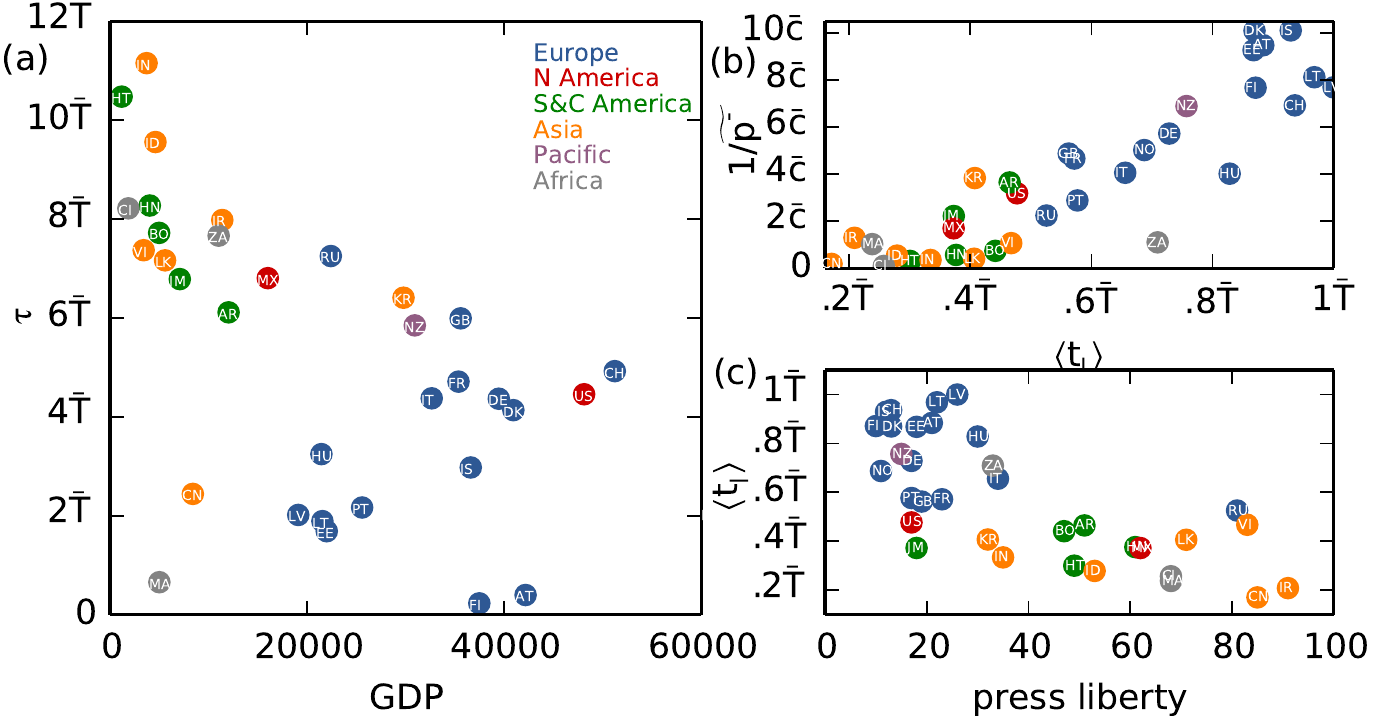}}
\caption{(Online version in colour.) Measures of innovation spreading vs. socioeconomic development in a country. (a) Predicted inverse speed of adoption spreading $\tau$ as a function of GDP per capita \cite{GDP} for various countries. (b) Theoretical inverse termination probability $1 / \widetilde{p^-}$ as a function of the empirical average account lifetime $\langle t_l \rangle$. (c) $\langle t_l \rangle$ as a function of a press liberty measure \cite{Liberty} (large scores imply weak liberties). $\bar{T}$ and $\bar{c}$ are arbitrary linear scaling constants with time dimensions for $\bar{T}$.}\label{fig:corrs}
\end{figure}

Our model may also be used to disclose relevant differences between the adoption dynamics of countries at various levels of societal and economical development. One characteristic indicator is the inverse speed of innovation diffusion, defined as the time $\tau$ when the theoretical $R_n(t)$ is maximal (see Supplementary Section S3.4). If we relate $\tau$ with one of the standard measures of economical development, GDP per capita \cite{GDP}, large differences emerge between countries (Fig. \ref{fig:corrs}.a). Specifically, the larger the GDP of a country, the faster the adoption process is in its society. Another way to characterize the adoption dynamics is through the average account lifetime, $\langle t_{l}\rangle=\langle t_t-t_a \rangle$, where $t_a$ and $t_t$ are the corresponding registration and termination times. We relate this empirical measure to its theoretical analogue, the inverse probability of termination $1/\widetilde{p^-}$ obtained from the fitted model process (Fig. \ref{fig:corrs}.b). Their correlation indicates that our model captures this dynamical property correctly. Moreover, the typical duration of user engagement uncovers clusters of countries at different levels of socio-economic development. This can be better understood by linking $\langle t_{l}\rangle$ with general civil liberty measures \cite{Liberty} (Fig. \ref{fig:corrs}.c). We observe that the weaker the press liberty is in a country, the shorter time online accounts are used there (other liberty measures in Supplementary Section S4.3). Such observations indicate a quantifiable dependence between the dynamics of innovation spreading and the socio-economic status of a country.

\section*{Discussion}

Our analysis of one of the largest online communication services worldwide aimed at clarifying several long-standing questions about the spreading mechanisms of novel technologies. We have shown that innovation diffusion can be interpreted as a competition between service adoption and termination; a process characterised, after a transition time, by constant rates and by a linearly increasing influence of user neighbours on service adoption. In addition we have integrated the identified mechanisms into a minimal modelling framework that provides accurate medium-term predictions for the spreading of an online service.

It should be pointed out that the present study has some limitations. First, the complete structure of society cannot be mapped by using online interactions only, as observations taken from any online social network underestimate the real number of contact peers of an ego. This incompleteness allows us only to estimate effective degrees and adoption thresholds. Second, the observation of correlated adoption does not necessarily imply the presence of actual social influence, only its possibility; even more so since other mechanisms like homophily cannot be synthesized from this dataset. Despite these limitations, the presented results provide strong evidence of key mechanisms driving the complex contagion of online technologies, up to a level of detail and scale that has not been possible before.

These results may help fill an enduring gap between the theoretical understanding and the empirical observation of social contagion phenomena, validating several earlier studies based on similar assumptions, like constant adoption rates and the effect of social influence. In addition, we have shown how the adoption of novel technologies is related to the societal and economical development of a country. Beyond the clear advantage of these observations for the design of marketing and business plans, they also provide further insight into the differences in the development of modern online societies.

\vspace{.1in}

{\bf Acknowledgments:} The authors gratefully acknowledge the support of M. Dumas and A. Saabas from STACC and Microsoft/Skype Labs, and from the ICTeCollective EU FP7 project. M.K. thanks R. Kikas for the data preparation and P. Gon\c{c}alves for useful comments. G.I. acknowledges the Academy of Finland for funding. J.K. thanks FiDiPro (TEKES) and the DATASIM EU FP7 project for support, and S. Fortunato for discussions. This research was partly funded by Microsoft/Skype Labs.

\begin{appendices}
\section{Model description}
For a static social network $G$ with degree distribution $\rho_k$, the probability that individual $i$ becomes a user is $p^+_i = p_a + p_p(1 - p_a)n_i$ with $n_i = N_i / k_i$. Here $N_i$ is the number of neighbours of $i$ that have already adopted the product and $k_i$ its degree. Furthermore, the probability that $i$ stops being a user is $p^-_i = p_r + p_s(1 - p_r)$. In the thermodynamic limit we assume that all agents with the same degree are statistically equivalent, allowing us to group individuals and write rate equations for each degree class $k$. We denote by $s_k$, $a_k$ and $r_k$ the average probabilities that a randomly chosen agent with degree $k$ is susceptible, adopter and removed, respectively. A first-order moment closure method leads to the rate equation $da_k / dt = \langle p^+_i \rangle s_k - \langle p^-_i \rangle a_k$. In other words, the average probability that an adopting agent becomes either removed or susceptible is $\langle p^-_i \rangle a_k = [p_r + p_s(1 - p_r)]a_k$, while the average probability that a susceptible individual adopts the product is $\langle p^+_i \rangle s_k = [p_a + p_p(1 - p_a)n_a]s_k$ with $n_a = \langle n_i \rangle$. This approximation ignores higher moments of the dynamical quantities $s_k$, $a_k$ and $r_k$, as well as any correlations between them. In the presence of degree-degree correlations in $G$ we have $n_a = \sum_{k'} (k' - 1) \rho_{k', k} a_{k'} / k'$, where $\rho_{k', k}$ is the conditional probability that an edge departing from an agent with degree $k$ arrives at an agent with degree $k'$. Similar rate equations can be written for $s_k$ and $r_k$, leading to a system of non-linear ordinary differential equations that determines adoption at the degree class level (for further details see Supplementary Section S3).
\end{appendices}

\pagebreak

{\LARGE \bf{Supplementary Information}}


\section{Data description}
\label{sec:materials}

This research is based on a dataset with a temporally detailed description of the social network of (anonymized) Skype users. It covers the history of users adopting Skype from September 2003 until March 2011 (2738 days), including registration events and contact network evolution for every registered users in Skype \footnote{At the end of 2010 Skype had more than 663 million registered accounts, as reported in \cite{SkypeIPO}.} all around the world. For each user, the dataset provides the following details:
\begin{itemize}
\item Date and country of registration.
\item Time-stamped events of link additions.
\end{itemize}

In the Skype network, when a user adds a friend to the contact list, the friend may confirm the contact invitation or not. Thus, links are added by means of the following events: contact addition and contact confirmation. In our study we capture trusted social links by retaining confirmed edges only, i.e. edges where both parties have acknowledged the connection. Failure to do so would lead to a mix of undesired and desired connections.

For our investigations we select user accounts with identified countries of registration and consider all confirmed connections between users. To receive the best estimation of node degrees in the underlying social network, we integrate the evolving Skype network for the whole available period and count the number of confirmed relationships per node, including international ties as well.

The adoption dynamics of a given country may be directly observed by assigning times of adoption ($t_a$) and termination ($t_t$) for all accounts. They are respectively defined as the dates of registration and last activity in Skype. Explicitly, we identify an account as terminated if its last activity happened earlier than one year prior to the end of the observation period. In this way we built the complete adoption and termination history of the Skype product for 2373 days.

\section{Empirical measures of spreading parameters}
\label{sec:params}

As discussed in the main text, we are able to estimate some of the model parameters directly from the empirical data. In what follows we describe in detail the definitions and limitations of the measurements for $p_a$, $p_p$, $p(n)$ and $p^-$, and shortly discuss the matching of the time scales of the real and model dynamics.

\subsection{Average number of social ties}
\label{sec:socTies}

Online social networks have the common limitation that, even while uncovering several characteristics of the real social graph underneath, they can only map a subset of the existing social relationships. This is simply because not everyone is registered to a given online service and thus not all social contacts are recorded. However, one may make the assumption that the online network, although only a sample of the real social graph, serves as a good proxy for the structure of society. This approximation is more reliable in technologically-advanced countries where the usage of online social services and communications is high, since the online sample is more representative.

We follow this line of thought and aggregate the evolving online social network of Skype between users of the same country for the whole available 7.5 years. In this way we receive a static aggregated structure as the best approximation of the actual social structure. Then we consider all international connections linking country users to accounts in any other nation, and finally estimate $\widetilde{ \langle k \rangle }$ as the average number of friends of a given individual, or ego. The observed ego networks are still incomplete, meaning that the estimated degree is bounded by its real value of the underlying social network, $\widetilde{ \langle k \rangle } \leq \langle k \rangle$. Even though this estimation process induces certain bias in the measurement (further discussed in section~\ref{sec:pressureAdopt}), its precision increases with larger values of $\widetilde{ \langle k \rangle }$.

\subsection{Probability of spontaneous adoption}
\label{sec:spontAdopt}

The probability of spontaneous adoption $p_a$ can be estimated directly from the data without any bias. The only things we need to know is when someone registered to Skype and whether such user was the first to adopt among his/her friends, independently of the degree of the user. Thus the probability of spontaneous adoption per unit time can be measured as,
\begin{equation}
\label{eq:empSpontAdopt}
p_a(t)=\dfrac{\# ad(t+\Delta t|SF=0)}{I-N_a(t)},
\end{equation}
where $\# ad(t+\Delta t|SF=0)$ is the number of users who adopted Skype during the period $[t,t+\Delta t]$, under the condition that none of their (later emerging) neighbors adopted before them. At each time step, this count is normalized by the total number of people who are not using Skype, i.e. the difference between the number of internet users $I$ in a given country and the number of users at time $t$, $N_a(t)$. By looking at the time evolution of $p_a(t)$ (orange curve in Fig.~2B, main text), it is clear that after an initial transient period this probability saturates and fluctuates around a constant value $\widetilde{p_a}$. Such value is the estimated average probability of spontaneous adoption in the interval $\Delta t$, which may then be used for the model calculations.

Even if the obtained rate approaches a constant for some countries, the observation time in the dataset is not sufficient to get the empirical estimate $\widetilde{p_a}$ in all cases. For consistency in our study, then, we leave $p_a$ as a free model parameter to be fitted. From the 34 countries studied, only 11 show a sufficiently fast adoption process such that $p_a(t)$ reaches a stationary value. A list of these countries is shown in Table~S\ref{table:1}, and the matter of an empirically vs. freely determined $p_a$ parameter is discussed in section \ref{sec:empfits}.

\subsection{Probability of peer-pressure adoption}
\label{sec:pressureAdopt}

Unfortunately, we cannot follow the same strategy in order to estimate the probability for peer-pressure adoption $p_p$. One could calculate a quantity analogous to that of Eq.~(\ref{eq:empSpontAdopt}),
\begin{equation}
\label{eq:empPeerAdopt}
p_p(t)=\dfrac{\# ad(t+\Delta t|SF\neq 0)}{I-N_a(t)},
\end{equation}
where $\# ad(t+\Delta t|SF\neq 0)$ is the number of adoption events per unit time, such that each adopting individual has at least one (or more) user neighbors at the moment of adoption. However, this quantity depends on the degree ($k_i$) and number of user friends ($N_i$) of any adopter $i$, and is thus driven by strong nonlinear network effects and node heterogeneity. This nonlinear behavior is evidenced by the time evolution of $p_p(t)$ (brown line in Fig.~2B, main text), where the calculated probability increases in time instead of saturating to a constant.

\begin{figure}
\centering
\subfigure[\hspace{76mm}]{\includegraphics[width=52mm]{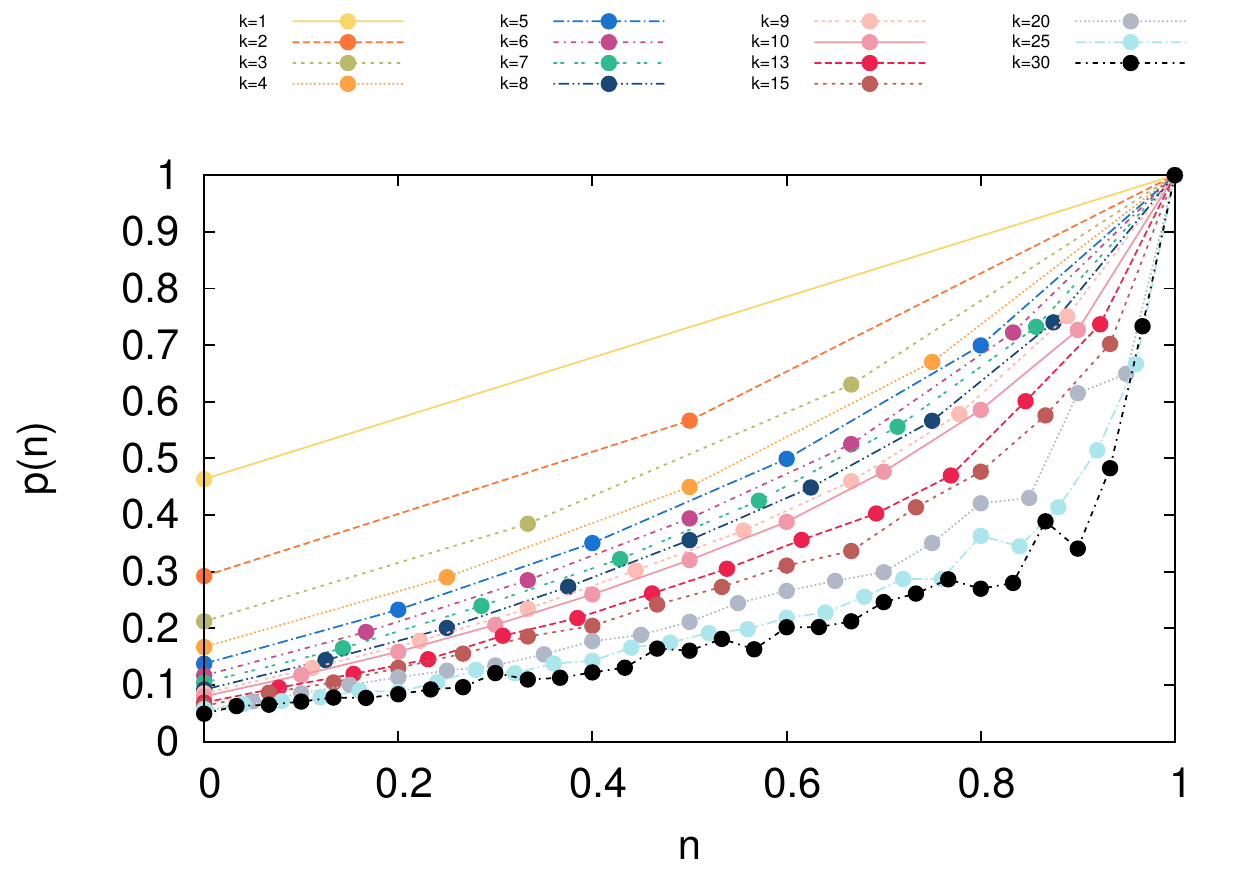}}
\subfigure[\hspace{76mm}]{\includegraphics[width=52mm]{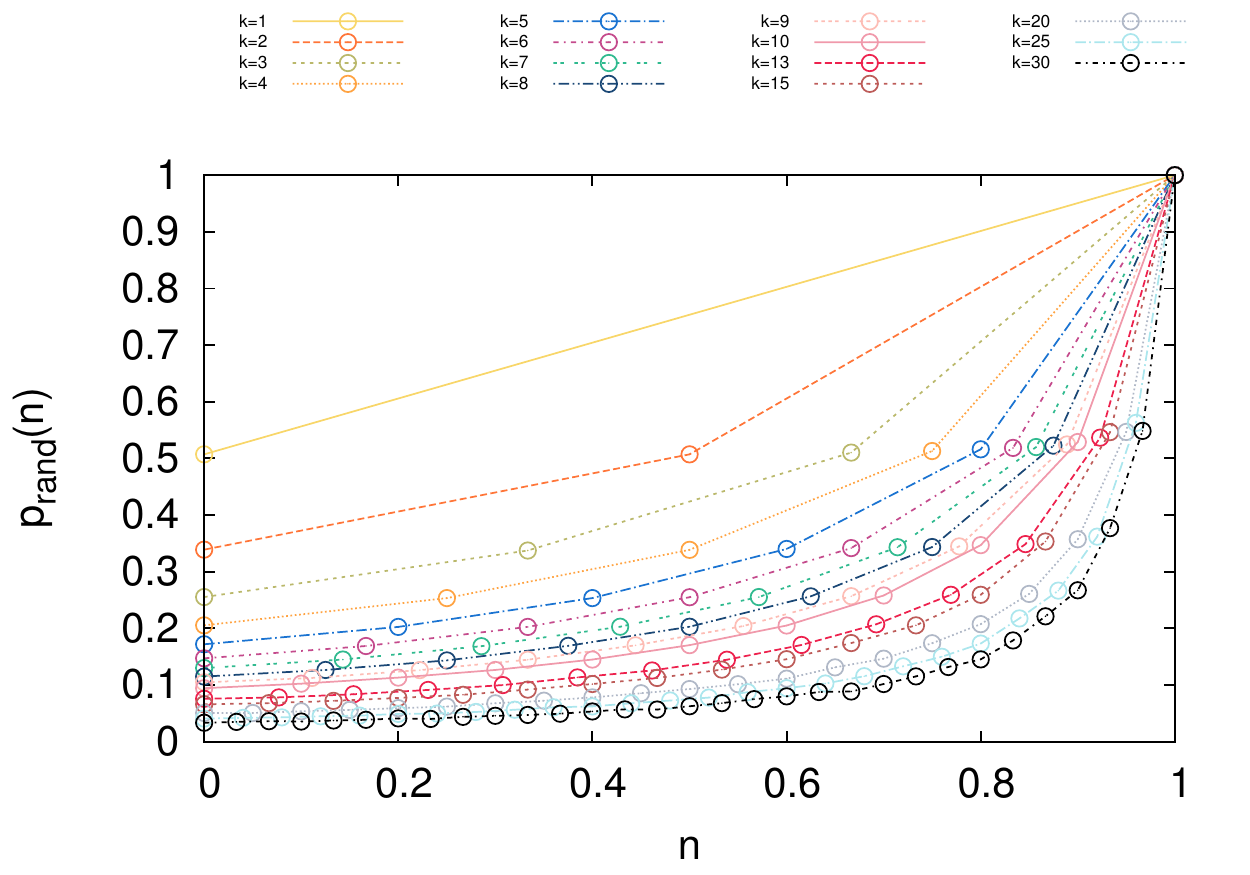}}
\subfigure[\hspace{76mm}]{\includegraphics[width=52mm]{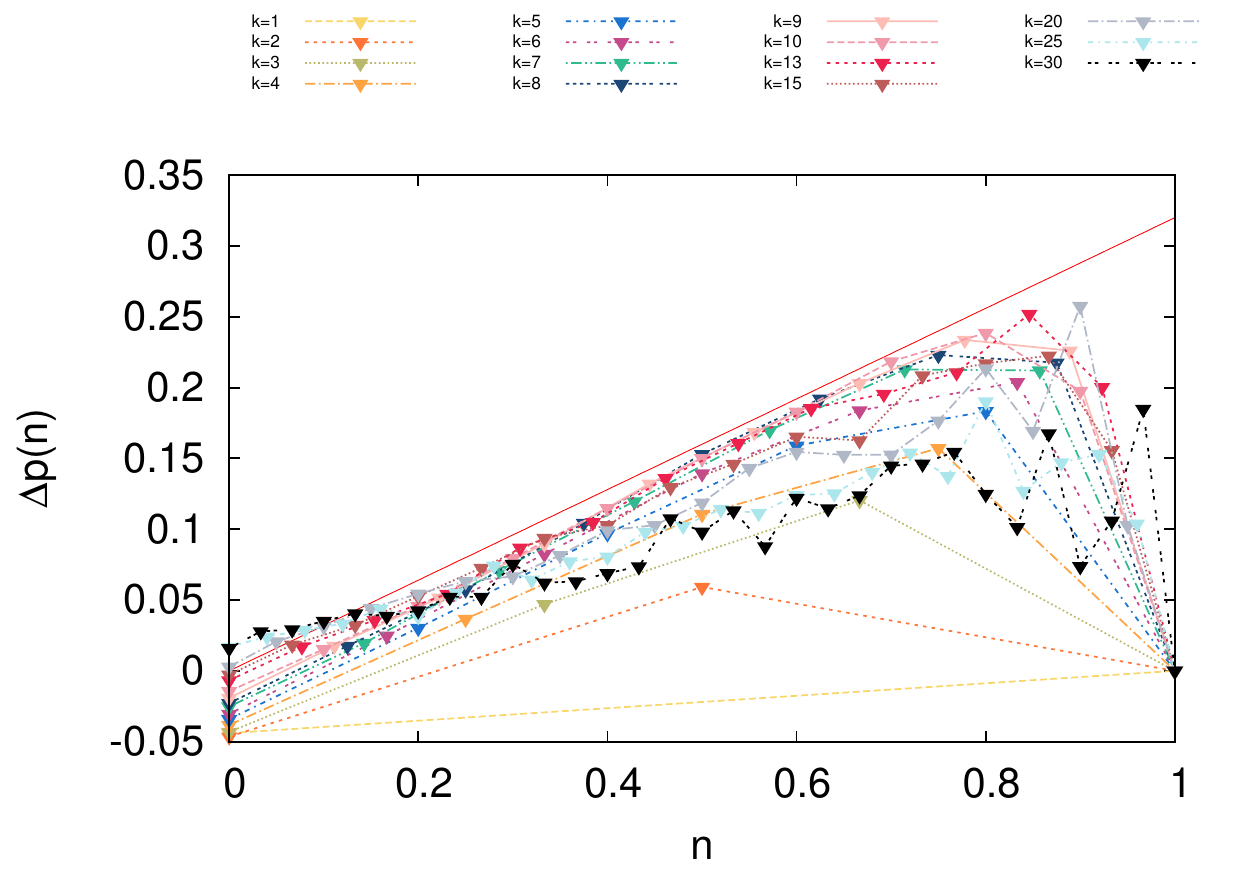}}
\caption{Probability of adoption driven by social influence. Conditional probability of adoption as a function of the effective fraction $\widetilde{n}$ of user neighbors. Panel (A) depicts $p(\widetilde{n}, \widetilde{k})$, the conditional probability for nodes of given effective degree $\widetilde{k}$, while panel (B) shows the similar measure $p_{\mathrm{rand}}(\widetilde{n})$ coming from a random reference system where the times of adoptions are shuffled. Panel (C) depicts the difference $\Delta p(\widetilde{n}, \widetilde{k})=p(\widetilde{n}, \widetilde{k})-p_{\mathrm{rand}}(\widetilde{n},\widetilde{k})$ calculated for each degree.}
\label{fig:CH3}
\end{figure}

A more appropriate way to quantify the effect of peer-pressure starts by measuring $n_i=N_i/k_i$, the fraction of user friends of each node $i$ at the time of adoption. However, in the integrated Skype graph we can only measure an effective degree $\widetilde{k_i}$ ($\leq k_i$) and thus an effective fraction $\widetilde{n_i}=N_i/\widetilde{k_i}$ ($\geq n_i$). Even as an approximation, this quantity can still show the qualitative effect of peer pressure in the likelihood of adoption.

With the value of $\widetilde{n_i}$ for all nodes in the integrated network, we can calculate the average conditional probability that a user adopts Skype given that an effective fraction $\widetilde{n}$ of his/her neighbours has already joined the network,
\begin{equation}
\label{eq:pn}
p(\widetilde{n})=\dfrac{\# ad(\widetilde{n})}{N-\sum_{m=0}^{m<\widetilde{n}}\# ad(m)},
\end{equation}
where the numerator is the number of users with a fraction $\widetilde{n}$ of Skype friends at the time of adoption. The denominator is the number of people with a larger or equal fraction $m \geq \widetilde{n}$, i.e. all individuals who had the chance to adopt Skype while having a fraction $\widetilde{n}$ of user neighbours. This count must exclude those people who have already adopted Skype due to weaker social pressure. Since we cannot see the entire social network (only the part uncovered by the Skype graph), this probability must take the extreme value $p(\widetilde{n})=1$ at $\widetilde{n}=1$ and is thus increasingly biased as $\widetilde{n} \rightarrow 1$. To see the real effect of peer pressure on the probability of adoption $p(\widetilde{n})$, we need to remove the effect of this bias. To this end, we define a random reference model where the termination time is set to infinity and shuffle the adoption times of all accounts. This null model is similarly biased by the effective quantities, but does not include the effects of social influence. Additionally, the null model conserves the adoption rates and keeps the social structure unchanged. In consequence, the empirical and null model values of the probability $p(\widetilde{n})$ differ only in the presence or absence of social influence, and thus their difference should quantify correctly the strength of social pressure and its effect on the probability of service adoption.

In Fig.~S\ref{fig:CH3}A we show $p(\widetilde{n},\widetilde{k})$, the conditional probability of adoption for nodes with given effective degree $\widetilde{k}$. Since $\widetilde{n}$ depends on $\widetilde{k}$ this probability takes discrete values, which is apparent for very small degrees. Similar effects can be observed in Fig.~S\ref{fig:CH3}B, where the corresponding curves $p_{\mathrm{rand}}(\widetilde{n},\widetilde{k})$ for the null model are shown for nodes of different degrees. The difference $\Delta p(\widetilde{n}, \widetilde{k}) = p(\widetilde{n}, \widetilde{k})-p_{\mathrm{rand}}(\widetilde{n},\widetilde{k})$ in Fig.~S\ref{fig:CH3}C shows that the effect of peer pressure increases linearly with the fraction of adopter neighbours, a rather robust behaviour in terms of degree. The slope of the linear regime may give an estimate $\widetilde{p_p}$ for the value of the probability of peer pressure adoption; however since we only use effective degrees in this calculation, the estimate is actually an upper limit for the real value, $p_p\leq \widetilde{p_p}$. The linear scaling breaks down around $\widetilde{n} \sim 0.8$, a value after which the peer pressure effect decreases radically. Several hypothesis can be introduced to explain this behaviour based on the individualist behaviour of an ego or on his or her reluctance to accept novel technologies, but these discussions are beyond the scope of the present study.

\subsection{Probability of termination}
\label{sec:churning}

As discussed in the main text, in our model the agents may terminate Skype usage in two different ways: either temporarily and going back into state $S$, or permanently going into state $R$, each with respective probability $p_s$ and $p_r$. In the former case agents may eventually readopt and enter state $A$ again, while in the latter they are removed and stay in state $R$ till the end of the process. Based on the empirical information provided by our dataset, we are not able to directly differentiate between temporary and permanent termination (since Skype accounts are not tied to uniquely identified individuals, who might or might not have multiple accounts) and thus measure the two probabilities independently. Instead we may measure the probability of overall termination,
\begin{equation}
\label{eq:stopProbEmp}
p^-(t) = \dfrac{\# tr(t+\Delta t)}{N_a(t)},
\end{equation}
where $\# tr(t+\Delta t)$ is the number of terminating users in the interval $[t,t+\Delta t]$, out of $N_a(t)$ possible users. Similarly to $p_a(t)$, the probability $p^-(t)$ is not biased by the incomplete social structure, and reaches a constant value $\widetilde{p^-}$ after an initial transient period (blue line in Fig.~2D, main text). This behaviour also holds for the decoupled probabilities of spontaneous and peer-pressure termination,
\begin{equation}
\label{eq:stopProbEmp2}
p_a^-(t) = \dfrac{\# tr(t+\Delta t,TF=0)}{N_a(t)} \hspace{.2in} \mbox{and} \hspace{.2in} p_p^-(t) = \dfrac{\# tr(t+\Delta t,TF\neq0)}{N_a(t)},
\end{equation}
where $TF$ is the number of terminated friends of the ego at the time of his/her own termination. Unlike in the case of adoption, both of these probabilities evolve towards a steady state, suggesting that termination is not driven by non-linear mechanisms and can be characterized by a constant rate. Note that by measuring $\widetilde{p^-}$ and using its definition in the model (see Eq.\ref{eq:stopProb}), we may estimate the parameter $p_s$ (or equivalently $p_r$) with the value,
\begin{equation}
\label{eq:psEmp}
\widetilde{p_s} = \frac{\widetilde{p^-}-p_r}{1-p_r}.
\end{equation}
Consequently, a steady termination process allows us to reduce the number of free parameters in the model by one.

\subsection{Time scales}
\label{sec:timeScales}

In order to match the time scales of the empirical and modelled rate sequences (seen in Fig. 2A of the main text), we let the time unit of the real process unchanged but introduce a constant $q = N_{int}/N_{pop}$ to rescale the model time as $t' = qt$. Here $N_{int}$ and $N_{pop}$ are the number of internet users and the population of a given country, respectively. Their ratio captures the average societal impact due to a digitally enabled sub-population that modulates the adoption of online products. Since we only rescale the model time while keeping the empirical sequences unchanged, this rescaling does not affect the time scale of the real adoption curves. As a result, the fits and model predictions are obtained in real time as well, independently of $q$. However, note that in order to correctly estimate model parameters from the data, we need to rescale their values as $p'_*(t')=qp_*(t)$ (where $p_*$ denotes either $p_a$ or $p^-$).

\section{Model}
\label{sec:model}

To simulate and predict the evolution of Skype adoption, we first need to synthesize the observed mechanisms into simple probabilistic rules, and then integrate them into a process modelling the interactions of a large number of individuals. We assume the existence of an underlying social network with arbitrary correlations and slow temporal evolution, in which individuals may choose to become users of the Skype product and give rise to an evolving account network. Our aim is to describe the temporal evolution of the account network with a suitable agent-based model. Under this approach we assume that individuals may start using Skype either by
\begin{description}
\item[(a)] adopting the product spontaneously, or by
\item[(b)] adopting the product due to peer pressure,
\end{description}
and terminate their use of the product either by
\begin{description}
\item[(c)] stopping usage temporarily with a chance of re-adoption, or by
\item[(d)] stopping usage permanently.
\end{description}

\subsection{Model description}
\label{sec:modDesc}

For a static social network $G$ of size $N$ and characterized by the adjacency matrix $\mathbf{A} = \{ a_{ij} \}$, the probability $p^+_i(t)$ that individual $i$ becomes a user at time $t$ is given by,
\begin{equation}
\label{eq:adoptProb}
p^+_i(t) = p_a + p_p(1 - p_a)n_i(t), \qquad n_i(t) = \frac{N_i(t)}{k_i},
\end{equation}
where $p_a \in [0, 1]$ is the probability of spontaneously adopting the product, $p_p \in [0, 1]$ is the probability to be affected by peer pressure, $N_i(t)$ is the number of neighbours of $i$ that at time $t$ have already adopted the product, and $k_i = \sum_j a_{ij}$ its degree. Since the density of product users in the neighbourhood of $i$ is $n_i(t)$, the time-dependent probability of adopting the product due to peer pressure is $p_p(1 - p_a)n_i(t)$. A peer pressure effect depending on the fraction of adopter neighbours rather than on their total number is reminiscent of other models of social activity, such as Watts' threshold model on global cascades \cite{Watts2002}.

Furthermore, the probability $p^-_i(t)$ that individual $i$ stops being a user at time $t$ is,
\begin{equation}
\label{eq:stopProb}
p^-_i(t) = p_r + p_s(1 - p_r),
\end{equation}
where $p_r, p_s \in [0, 1]$ are the probabilities of halting usage either permanently or temporarily. Overall, agents can be classified into sets of susceptible ($S$), adopting ($A$) and removed ($R$) individuals, describing respectively people who may adopt the product later, are users already, and will never use it again. The flow $S \to A$ is regulated by $p_a$ and $p_p$, $A \to R$ by $p_r$, and $A \to S$ by $p_s$.

In the thermodynamic limit $N \to \infty$, the process of adoption at the user level can be well characterized with a master equation formalism. We assume that all agents with the same degree are statistically equivalent, allowing us to group individuals and write rate equations for each degree class $k$ \cite{Barrat2008}. We denote by $s_k(t)$, $a_k(t)$ and $r_k(t)$ the average probabilities that a randomly chosen agent with degree $k$ is susceptible, adopter and removed, respectively. In the limit of large system size these probabilities are equal to the densities of susceptible, adopting and removed individuals in the degree class $k$, so that $s_k + a_k + r_k = 1$ $\forall\, t, k$.

Let us denote by $\rho_k$ the degree distribution of the static social network $G$, that is, the probability that a randomly chosen agent has degree $k$. With it the average probability of an individual belonging to the sets $S$, $A$ and $R$ are correspondingly given by
\begin{equation}
\label{eq:averDens}
s(t) = \sum_k \rho_k s_k(t), \qquad  a(t) = \sum_k \rho_k a_k(t), \qquad r(t) = \sum_k \rho_k r_k(t),
\end{equation}
with the normalization condition $s + a + r = 1$ $\forall\, t$. Our task is to find rate equations for the probabilities $s_k$, $a_k$ and $r_k$ that correspond to the dynamics set by Eqs.~(\ref{eq:adoptProb}) and~(\ref{eq:stopProb}), solve them and average over the distribution $\rho_k$ to get the time dependence of the probabilities $s$, $a$ and $r$ in Eq.~(\ref{eq:averDens}), so as to describe the evolution of the account network at a global level.

Through a simple first-order moment closure method, the rate equation for $a_k$ can be written as $da_k / dt = \langle p^+_i \rangle s_k - \langle p^-_i \rangle a_k$. In other words, the average probability that an adopting agent with degree $k$ becomes either removed or susceptible is $\langle p^-_i \rangle a_k = [p_r + p_s(1 - p_r)]a_k$, while the average probability that a susceptible individual in the degree class $k$ adopts the product is $\langle p^+_i \rangle s_k = [p_a + p_p(1 - p_a)n_a]s_k$, where $n_a(t) = \langle n_i(t) \rangle$ is the average probability that the neighbour of a susceptible agent has adopted already. This approximation ignores higher moments of the dynamical quantities $s_k$, $a_k$ and $r_k$, as well as any correlations between them. In the presence of degree-degree correlations in $G$,
\begin{equation}
\label{eq:adoptNeigh}
n_a = \sum_{k'} \frac{k' - 1}{k'} \rho_{k', k} a_{k'},
\end{equation}
with $\rho_{k', k}$ the conditional probability that an edge departing from an agent with degree $k$ arrives at an agent with degree $k'$ \cite{Barrat2008}. We can write similar equations for $s_k$ and $r_k$ to arrive at the system,
\begin{eqnarray}
\label{eq:prodODEs1}
\frac{da_k}{dt} & = &  [p_a + p_p(1 - p_a)n_a]s_k - [p_r + p_s(1 - p_r)]a_k \\
\label{eq:prodODEs2}
\frac{ds_k}{dt} & = & -[p_a + p_p(1 - p_a)n_a]s_k + p_s(1 - p_r)a_k \\
\label{eq:prodODEs3}
\frac{dr_k}{dt} & = & p_r a_k
\end{eqnarray}
that forms a system of non-linear ordinary differential equations determining the adoption dynamics at the degree class level.

\subsection{Uncorrelated random networks}
\label{sec:uncorNets}

In order to progress further we need to simplify Eqs.~(\ref{eq:prodODEs1})-(\ref{eq:prodODEs3}) by making additional assumptions about the average probability $n_a$. If the social network $G$ is considered as an uncorrelated random network, the conditional probability $\rho_{k', k}$ does not depend on $k$ any more and it takes the simple form $\rho_{k', k} = k' \rho_{k'} / \langle k \rangle$, where $\langle k \rangle = \sum_k k \rho_k$ is the average degree in $G$. By substituting it into Eq.~(\ref{eq:adoptNeigh}) we obtain,
\begin{equation}
\label{eq:uncorrAdopt}
n_a = \frac{1}{\langle k \rangle} \sum_k (k - 1) \rho_k a_k.
\end{equation}
Since the sum in $n_a$ goes through all values of $k$, Eqs.~(\ref{eq:prodODEs1})-(\ref{eq:prodODEs3}) for all degree classes will be identical apart from initial conditions. In our case it is relevant to consider $s^0_k = 1$ and $a^0_k = r^0_k = 0$ $\forall\, k$, which further simplifies the dynamics and gives $a_k = a$ $\forall\, k$, that is, $n_a = a (\langle k \rangle - 1) / \langle k \rangle$.

Moreover, since $\rho_k$ is not a function of $t$, we can take an average by using Eq.~(\ref{eq:averDens}) and write,
\begin{eqnarray}
\label{eq:uncorrODEs1}
\frac{da}{dt} & = & [p_a + p_{pk}(1 - p_a)a]s - [p_r + p_s(1 - p_r)]a \\
\label{eq:uncorrODEs2}
\frac{ds}{dt} & = & -[p_a + p_{pk}(1 - p_a)a]s + p_s(1 - p_r)a
\end{eqnarray}
now a planar system of non-linear ordinary differential equations that describes the evolution of the account network at a global level, where we have redefined the 
peer-pressure influence as the effective probability,
\begin{equation}
\label{eq:peerDeg}
p_{pk} = \frac{\langle k \rangle - 1}{\langle k \rangle} p_p.
\end{equation}
For large values of $\langle k \rangle$, the effect of the social network's degree is minimal and $p_{pk} \sim p_p$.

\begin{figure}
\centering
\subfigure[\hspace{75mm}]{\includegraphics[width=75mm]{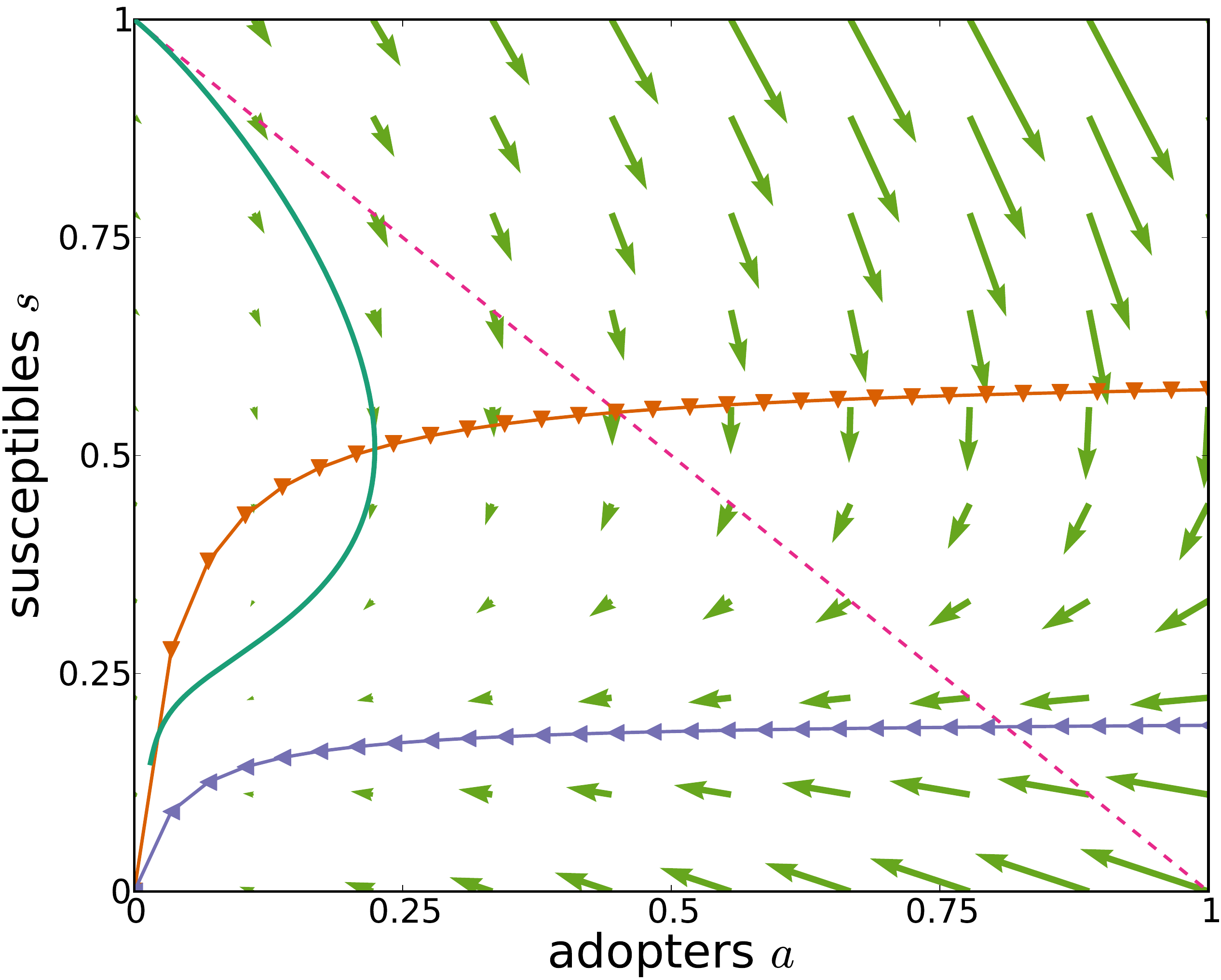}}
\subfigure[\hspace{75mm}]{\includegraphics[width=75mm]{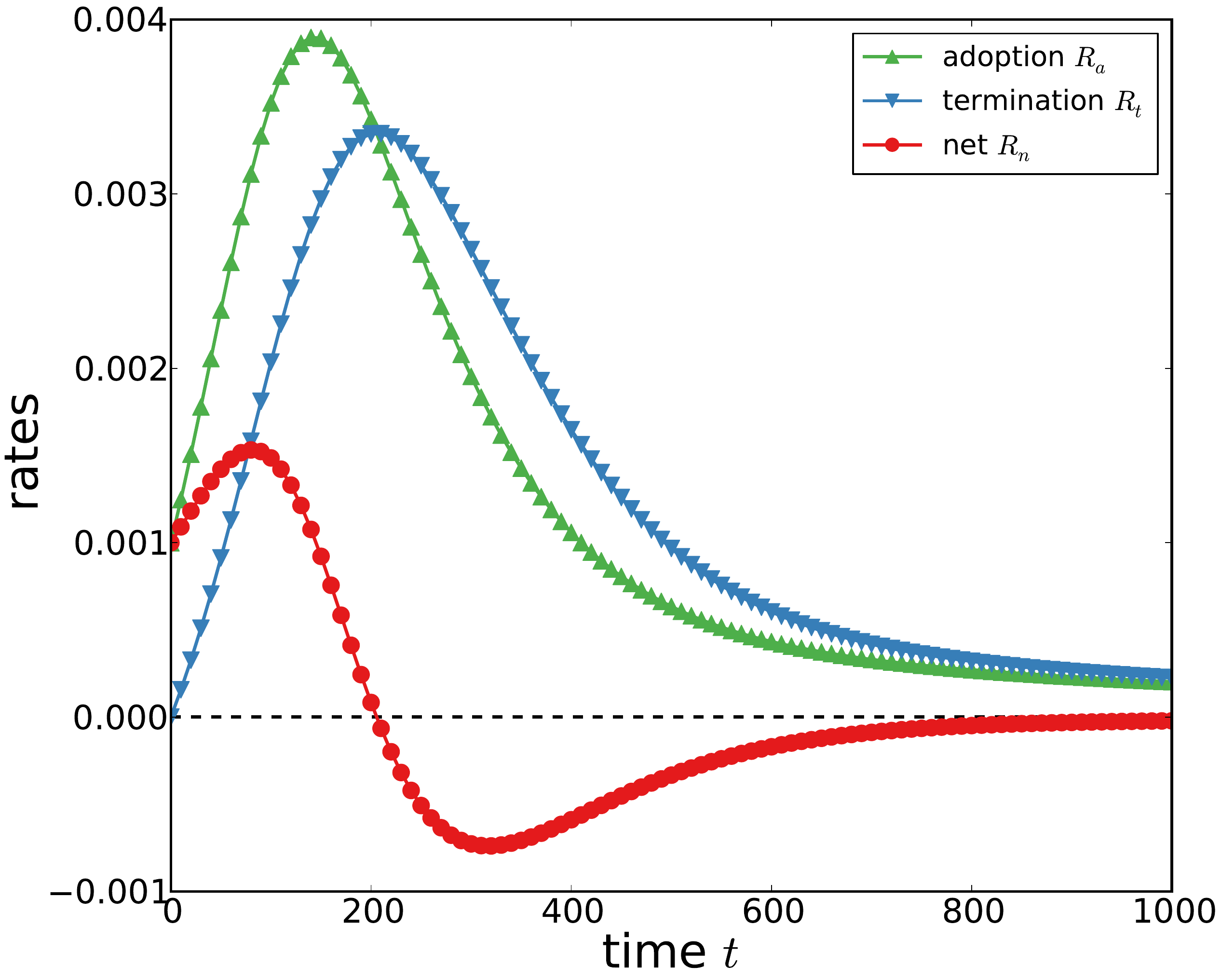}}\\
\caption{Numerical solution of the model process. (A) Phase portrait of the dynamics set by Eqs.~(\ref{eq:uncorrODEs1})-(\ref{eq:uncorrODEs2}). The arrows represent the flow of the dynamical system, symbols show the 0-clines $s_a$ ($\triangledown$) and $s_s$ ($\triangleleft$), and the dashed line is the upper limit for the allowed phase space $(a, s)$ according to the condition $s + a + r = 1$. The continuous line is an example trajectory with initial condition $(a^0, s^0) = (0, 1)$ and parameters $p_a = 0.001$, $p_p = 0.05$, $p_s = 0.005$, $p_r = 0.01$ and $\langle k \rangle = 2$. (B) Time evolution of the rates of adoption ($R_a$), termination ($R_t$) and net change ($R_n = R_a - R_t$) for the same set of parameters.}
\label{fig:uncorrDynam}
\end{figure}

The non-linear system in Eqs.~(\ref{eq:uncorrODEs1})-(\ref{eq:uncorrODEs2}) can be approached qualitatively by a linear stability analysis. The 0-clines of the system (i.e. the curves at which $da / dt$ and $ds / dt$ are respectively zero) can be written as,
\begin{equation}
\label{eq:0clines}
s_a = \frac{[p_r + p_s(1 - p_r)]a}{p_a + p_{pk}(1 - p_a)a}, \qquad s_s = \frac{p_s(1 - p_r)a}{p_a + p_{pk}(1 - p_a)a},
\end{equation}
so that $s_a - s_s = p_r a / [p_a + p_{pk}(1 - p_a)a]$, and for $p_r \neq 0$ the only fixed point of the dynamics is $(a^{\infty}, s^{\infty}) = (0, 0)$. Fig.~S\ref{fig:uncorrDynam}A shows the flow of the system~(\ref{eq:uncorrODEs1})-(\ref{eq:uncorrODEs2}) and the 0-clines according to Eq.~(\ref{eq:0clines}).

By denoting $da / dt = f_a(a, s)$ and $ds / dt = f_s(a, s)$, the dynamics near the fixed point is determined by the Jacobian matrix,
\begin{equation}
\label{eq:jacobMat}
J(a, s) =
\begin{pmatrix}
df_a / da & df_a / ds \\
df_s / da & df_s / ds
\end{pmatrix} =
\begin{pmatrix}
p_{pk}(1 - p_a)s - p_r - p_s(1 - p_r) & p_a + p_{pk}(1 - p_a)a \\
-p_{pk}(1 - p_a)s + p_s(1 - p_r) & -p_a -p_{pk}(1 - p_a)a
\end{pmatrix},
\end{equation}
i.e. $(da / dt, ds / dt) = J(0, 0) (a, s)$ for $(a, s) \sim (0, 0)$, and the stability of the fixed point is given by the eigenvalues $\lambda_{\pm}$ of $J(0, 0)$,
\begin{equation}
\label{eq:eigenSyst}
0 =
\begin{vmatrix}
-p_r - p_s(1 - p_r) - \lambda_{\pm} & p_a \\
p_s(1 - p_r) & -p_a - \lambda_{\pm}
\end{vmatrix} \, \implies \,
\lambda_{\pm} = \frac{1}{2} \left( \tau \pm \sqrt{\tau^2 - 4\Delta} \right),
\end{equation}
where $\tau = -[p_a + p_r + p_s(1 - p_r)] < 0$ and $\Delta = p_a p_r > 0$, as long as $p_a, p_r, p_s \neq 0$. Since $\tau < 0$ and $\tau^2 - 4\Delta > 0$ the fixed point is a stable node, meaning that $(a^{\infty}, s^{\infty})$ is indeed the final state of the dynamics and attracts all trajectories of the phase space. Fig.~S\ref{fig:uncorrDynam}A shows an example trajectory starting from $(a^0, s^0) = (0, 1)$ for given parameter values, progressively approaching the fixed point.

Finally we consider the rates at which individuals adopt the product [$R_a$(t)] and stop using it [$R_t(t)$], as well as the rate of effective or net change $R_n(t) = R_a(t) - R_t(t)$, since they can be directly compared with the empirical data. By construction these are equal to the gain and loss terms in the rate equation for $a$, that is,
\begin{equation}
\label{eq:ratesDef}
R_a(t) = [p_a + p_{pk}(1 - p_a)a]s, \qquad R_t(t) = [p_r + p_s(1 - p_r)]a.
\end{equation}
A numerical evaluation of the rates $R_a(t)$, $R_t(t)$ and $R_n(t)$ for given parameter values is shown in Fig.~S\ref{fig:uncorrDynam}B.

\subsection{Numerical simulations}
\label{sec:numerSimul}

\begin{figure}
\centering
\subfigure[\hspace{20mm}$R_a(t)$\hspace{20mm}]{\includegraphics[width=50mm]{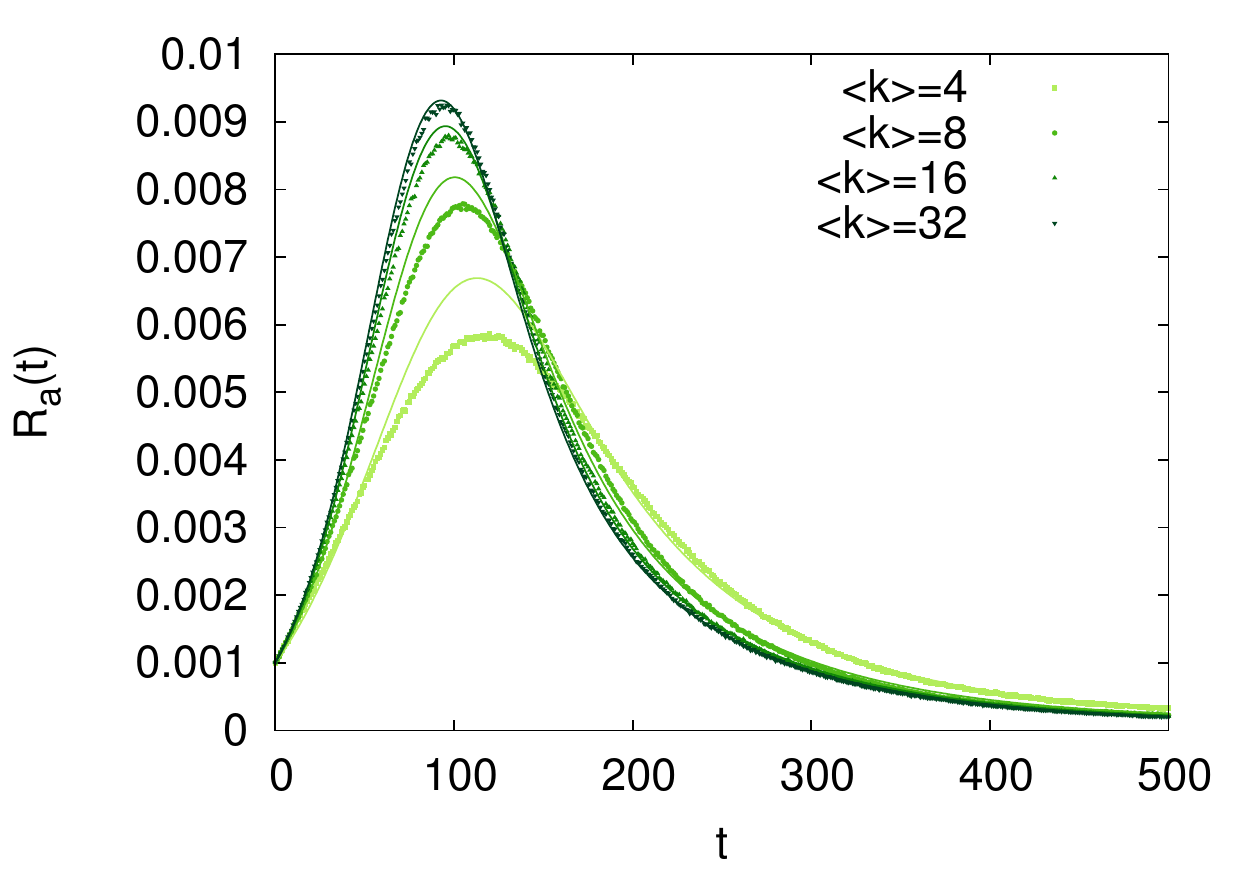}}
\subfigure[\hspace{20mm}$R_t(t)$\hspace{20mm}]{\includegraphics[width=50mm]{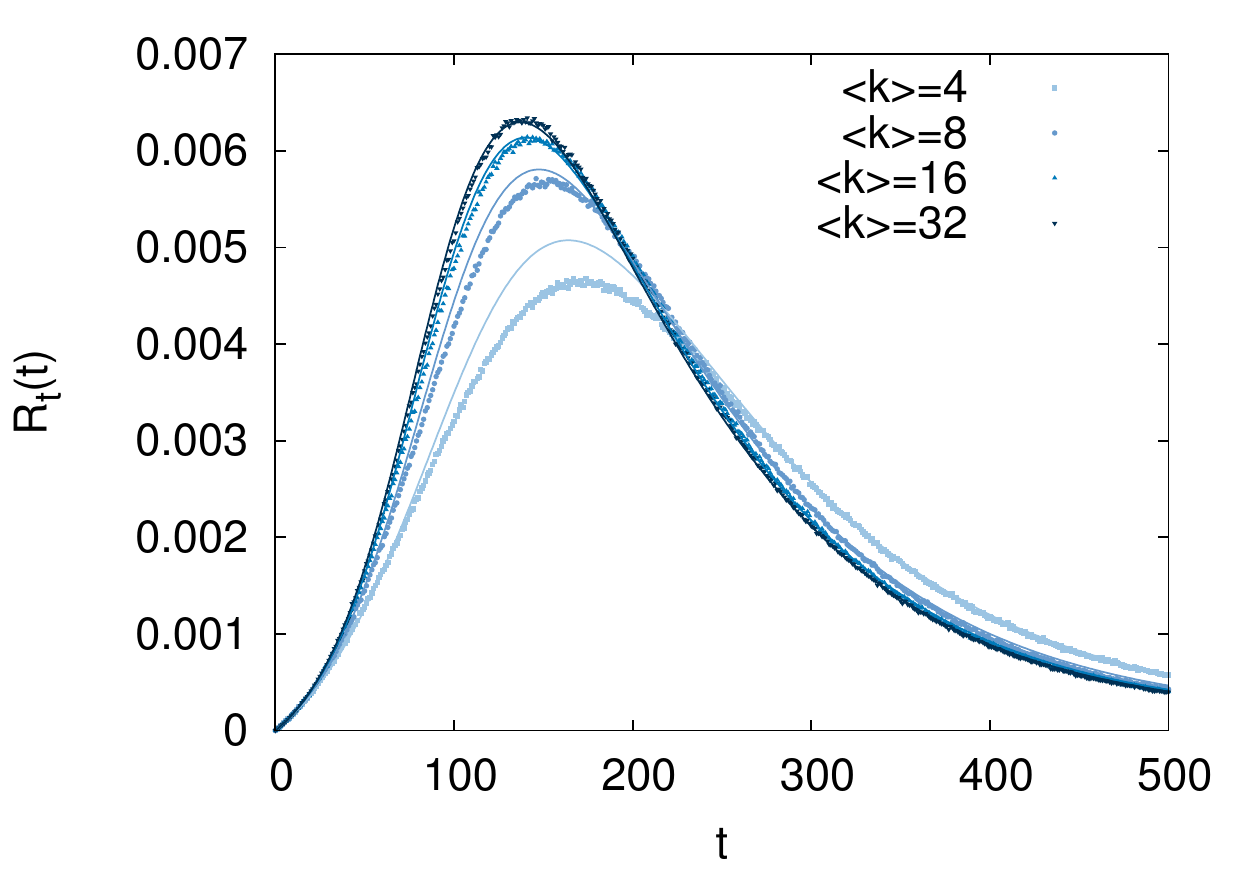}}
\subfigure[\hspace{20mm}$R_n(t)$\hspace{20mm}]{\includegraphics[width=50mm]{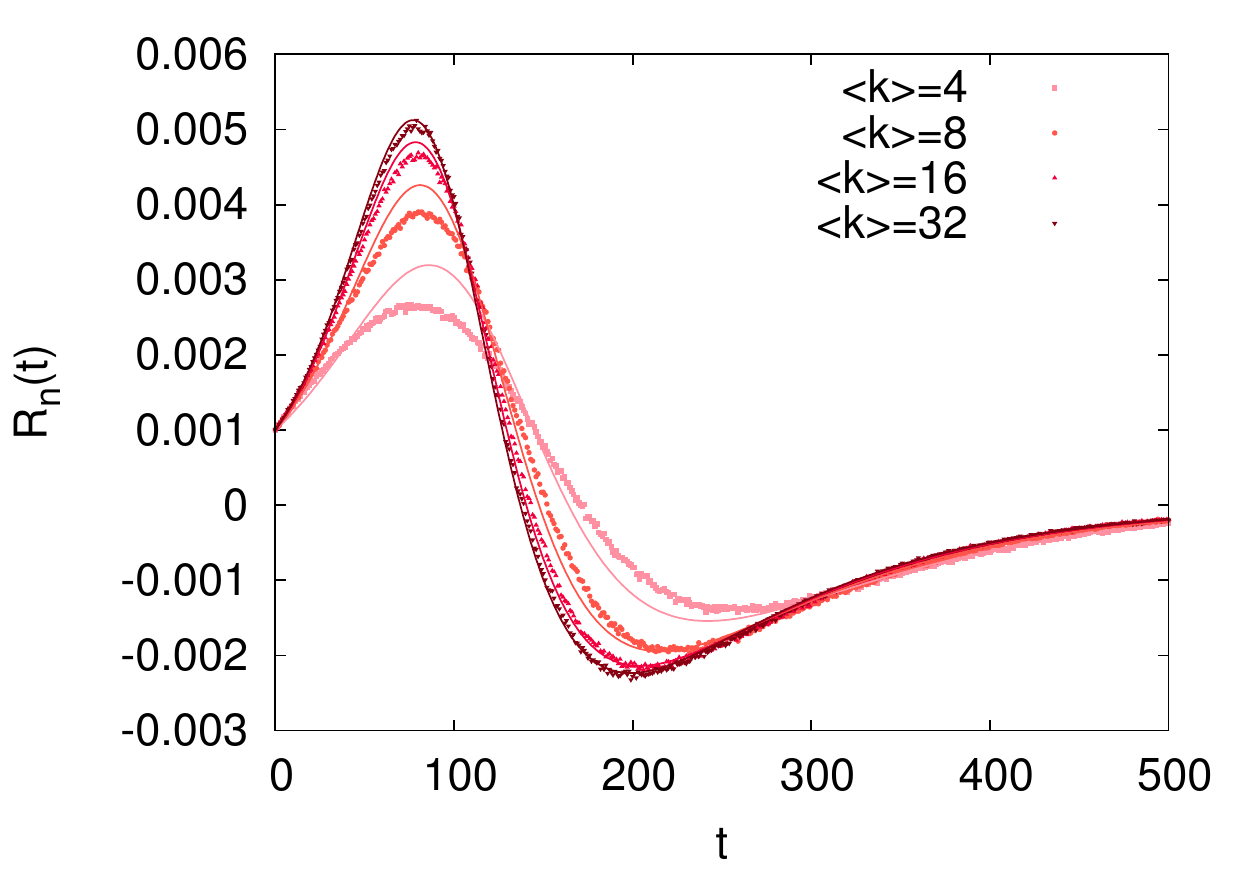}}\\
\caption{Numerical simulations of the adoption process on SF networks with different average degrees. We use networks with average degree $\langle k \rangle=4$ ($\Box$), $8$ ($\circ$), $16$ ($\triangle$), $32$ ($\triangledown$) and size $N=10^5$. We show (A) the rate of adoption $R_a(t)$ (green), (B) the rate of termination $R_t(t)$ (blue), and (C) the rate of net change $R_n(t) = R_a(t)-R_t(t)$ (red). Symbols indicate averages of numerical simulations, while solid lines are curves derived from Eq.~(\ref{eq:ratesDef}) with parameters $p_a = 0.001$, $p_p = 0.05$, $p_s = 0.005$ and $p_r = 0.01$.}
\label{fig:ns1}
\end{figure}

To verify our theoretical considerations we can implement the previous agent-based model and compare the solutions of Eq.~(\ref{eq:ratesDef}) with large-scale numerical simulations on synthetic network structures. The Skype network is strongly heterogeneous and presents a power-law like degree distribution as many other social networks. To get a similar structure for our numerical simulations we implement the Barab{\'a}si-Albert model \cite{Barrat2008} and generate scale-free (SF) networks with a similar exponent, uncorrelated degrees and scalable average degree. We use this topology as a model for the background social network and on the top of it we perform the process defined by Eqs.~(\ref{eq:adoptProb})-(\ref{eq:stopProb}).

During our theoretical considerations we have taken a mean-field approach that provides better accuracy if $N \rightarrow \infty$ and the average degree $\langle k \rangle$ of the background network is large. The validity of such approximation (and the correctness of our solution) can be verified by comparing the characteristic rates calculated from averages of large-scale simulations to the rates provided by Eq.~(\ref{eq:ratesDef}), while using the same parameter values. In Fig.~S\ref{fig:ns1} we show numerical simulations for synthetic networks with different average degrees, averaged over 100 realizations of the adoption process. As $\langle k \rangle$ increases the discrepancy between the theoretical and simulated rates reduces considerably, until finally in the limit of large $\langle k \rangle$ the fit between both rates is very accurate, validating our theoretical solution of the model.

\begin{figure}
\centering
\subfigure[\hspace{20mm}$R_a(t)$\hspace{20mm}]{\includegraphics[width=50mm]{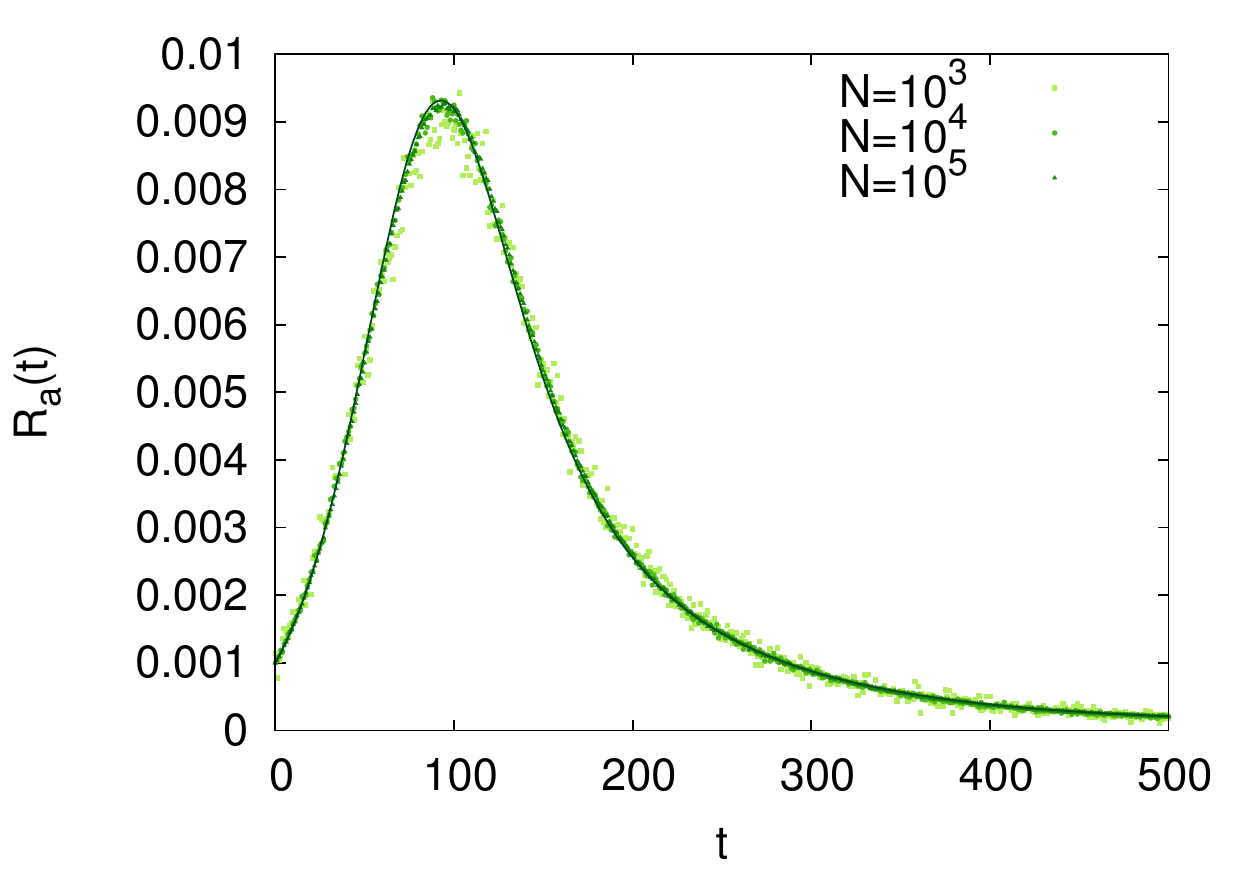}}
\subfigure[\hspace{20mm}$R_t(t)$\hspace{20mm}]{\includegraphics[width=50mm]{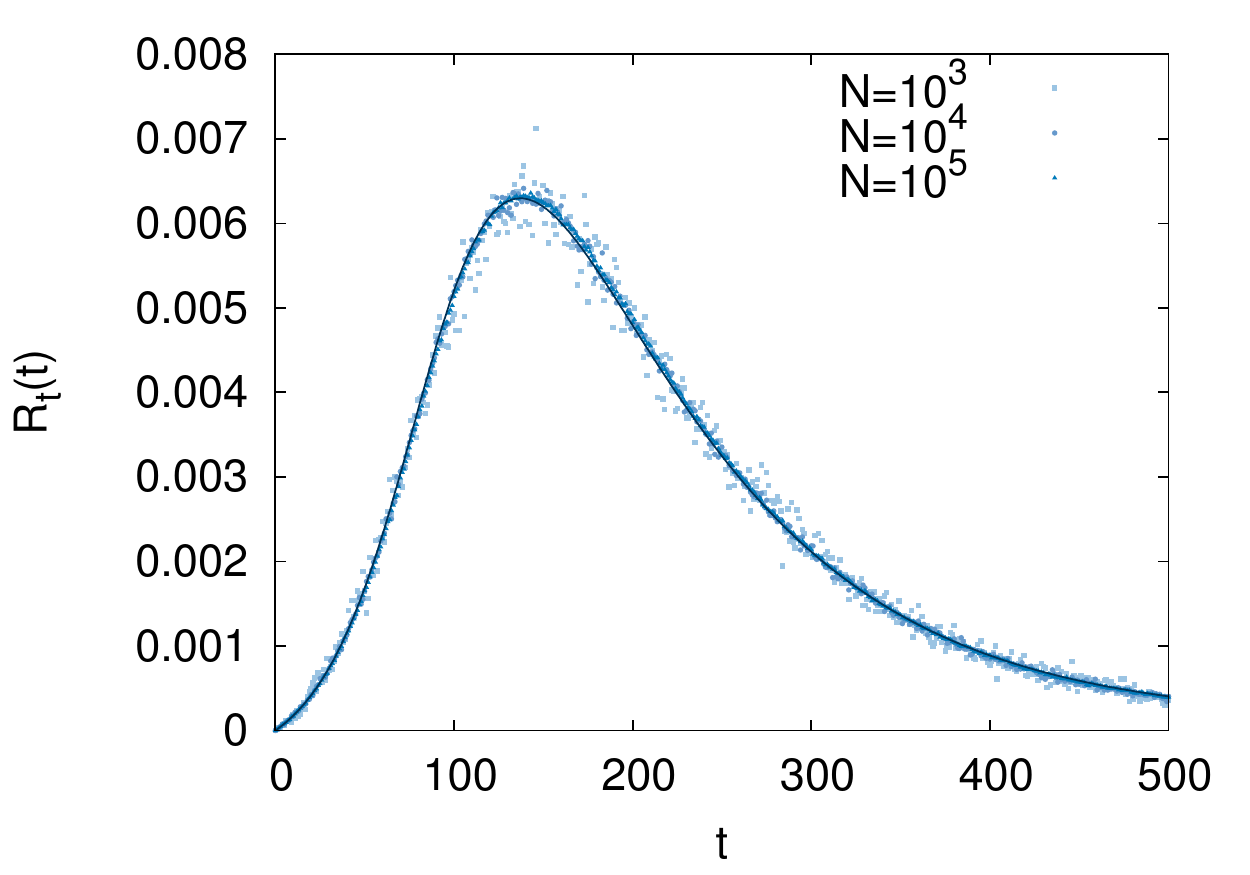}}
\subfigure[\hspace{20mm}$R_n(t)$\hspace{20mm}]{\includegraphics[width=50mm]{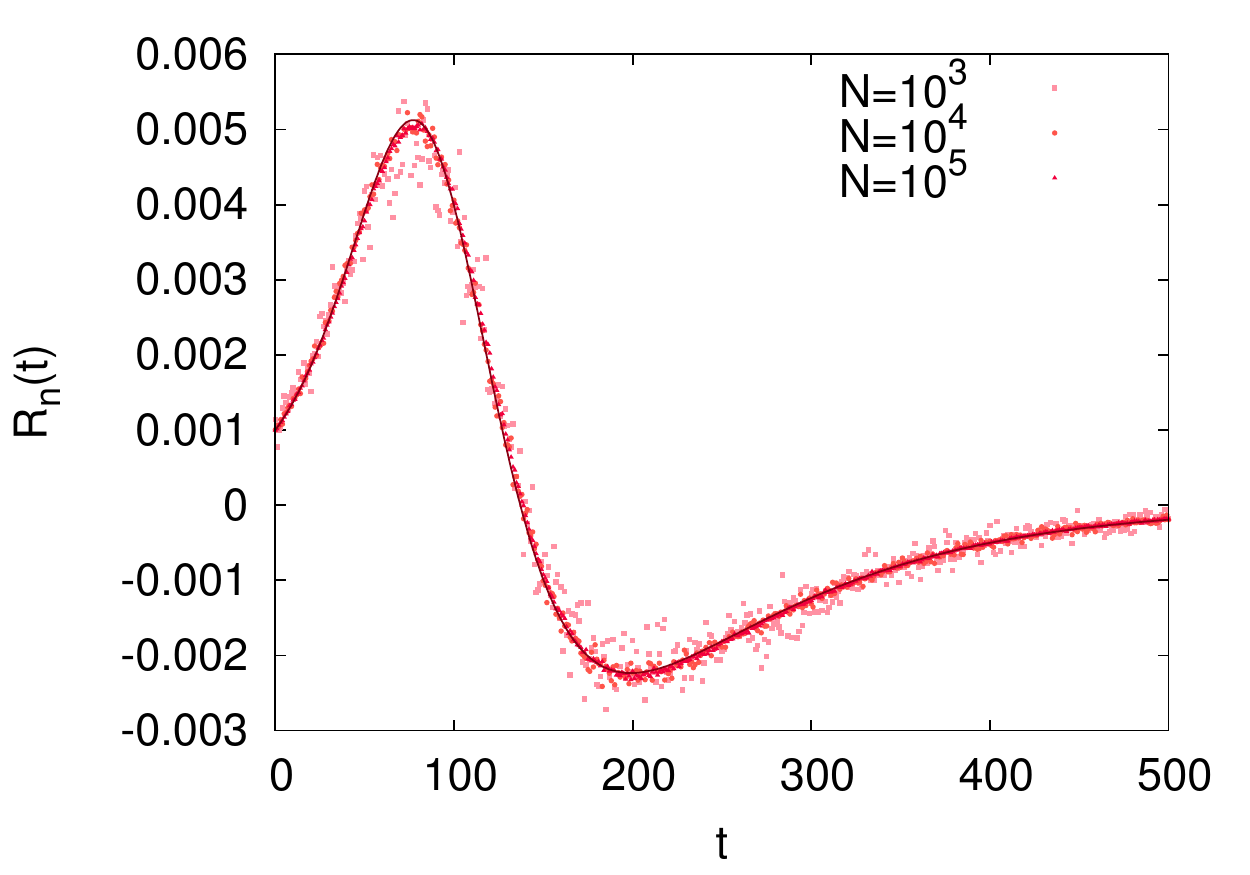}}\\
\caption{Numerical simulations of the adoption process on SF networks with different sizes. We use networks with size $N = 10^3$ ($\Box$), $10^4$ ($\circ$), $10^5$ ($\triangle$) and average degree $\langle k \rangle = 32$. We show (A) the rate of adoption $R_a(t)$ (green), (B) the rate of termination $R_t(t)$ (blue), and (C) the rate of net change $R_n(t)$ (red). Symbols indicate averages of numerical simulations, while solid lines are curves derived from Eq.~(\ref{eq:ratesDef}) with the same parameter values as in Fig.~S\ref{fig:ns1}.}
\label{fig:ns2}
\end{figure}

The solution given in Eq.~(\ref{eq:ratesDef}) suggests no system size dependence of the normalized characteristic rates. This can be confirmed by simulations where only the size $N$ of the static network is varied. The averages of $100$ realizations shown in Fig.~S\ref{fig:ns2} indicate that even if in smaller networks the simulated rates have larger deviation, their averages overlap with the theoretical curve.

\subsection{Spreading scenarios}
\label{sec:spreadScen}

\begin{figure}
\centering
\subfigure[\hspace{30mm}Rates\hspace{30mm}]{\includegraphics[width=70mm]{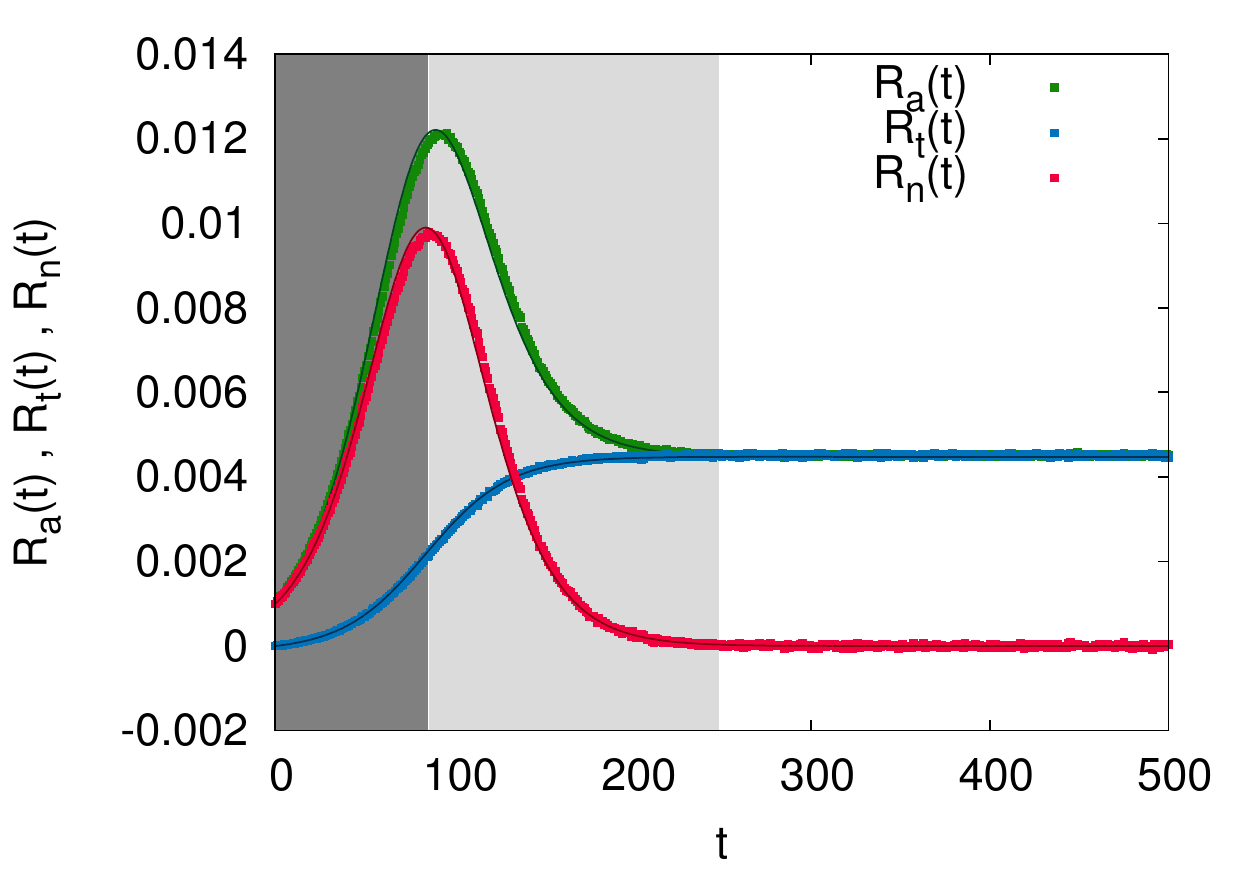}}
\subfigure[\hspace{22mm}Cumulative rates\hspace{22mm}]{\includegraphics[width=70mm]{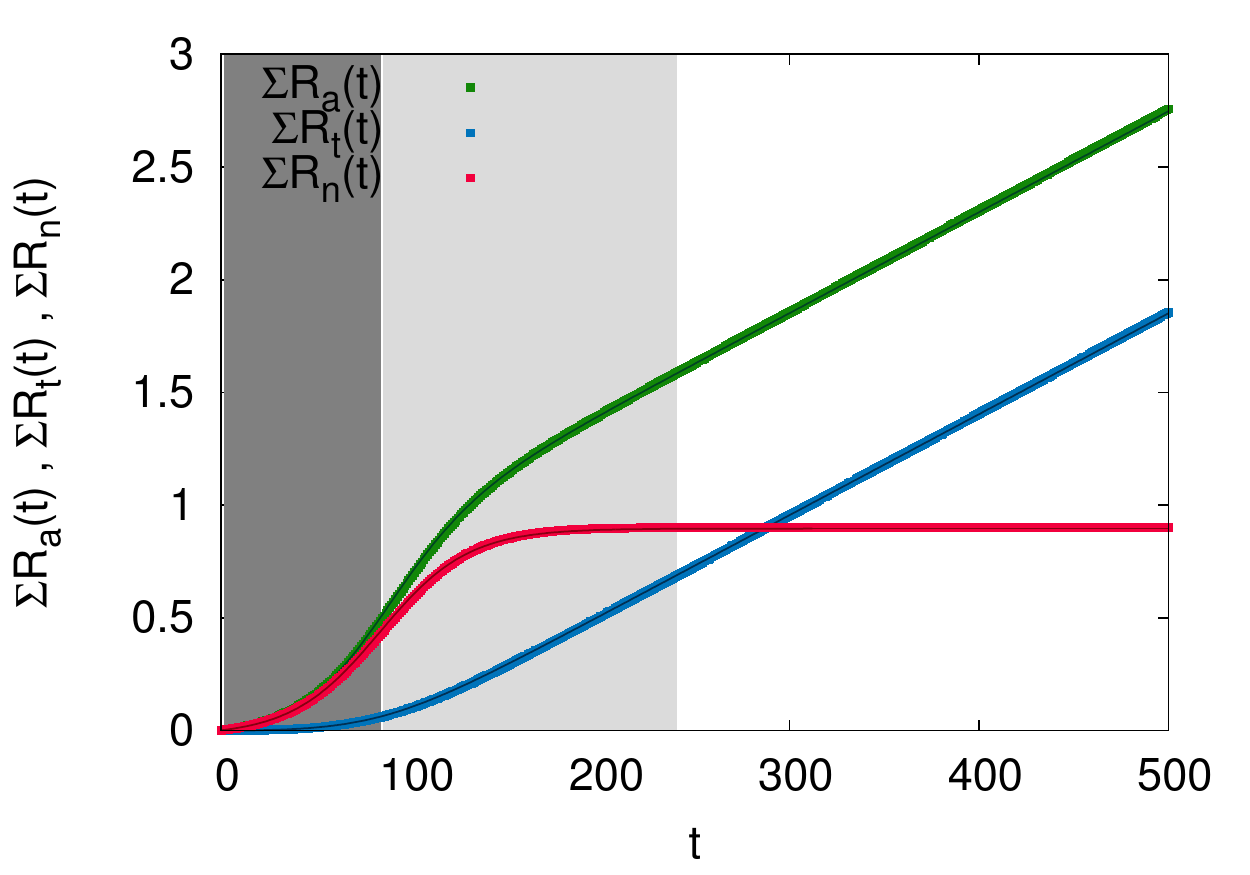}}\\
\subfigure[\hspace{30mm}Rates\hspace{30mm}]{\includegraphics[width=70mm]{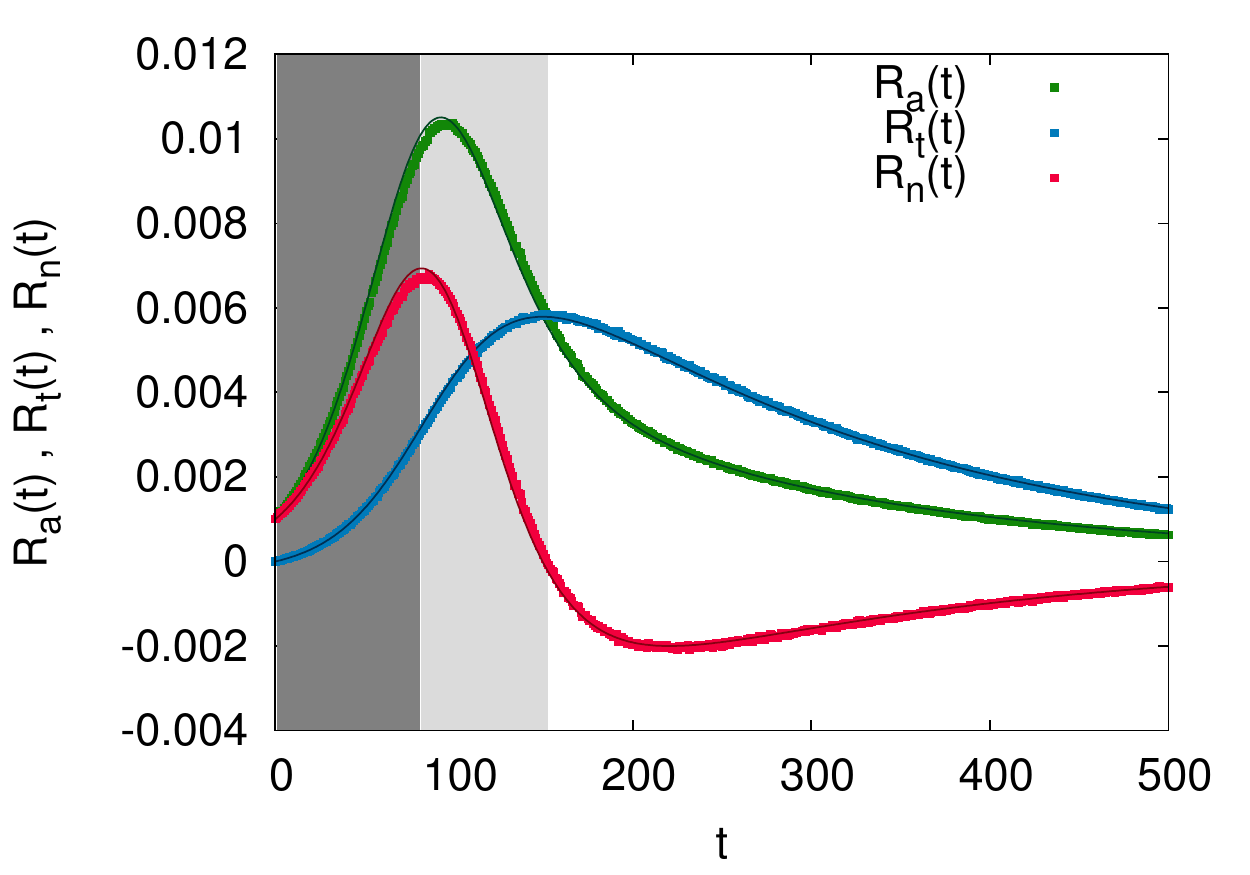}}
\subfigure[\hspace{22mm}Cumulative rates\hspace{22mm}]{\includegraphics[width=70mm]{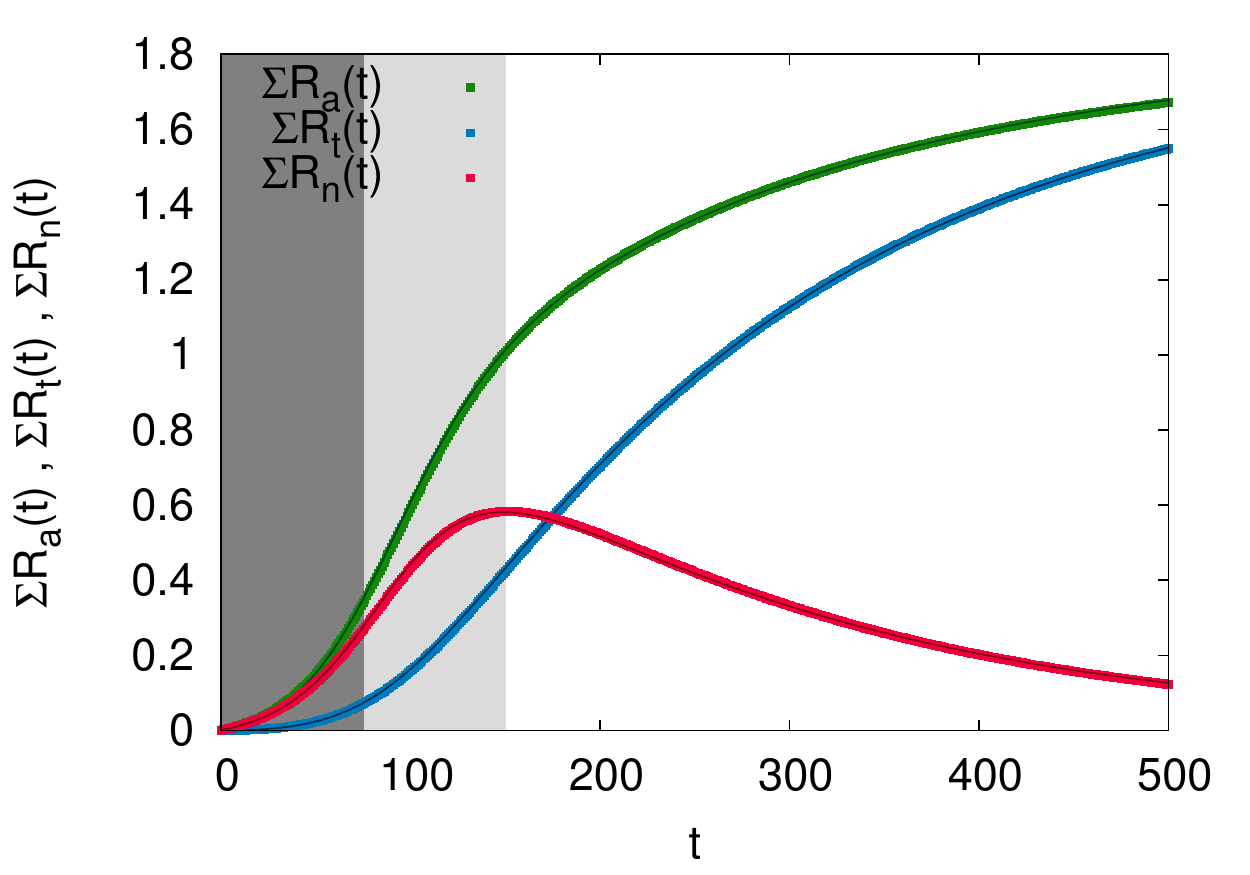}}\\
\caption{Numerical simulations and theoretical prediction of adoption processes. Processes were run with parameters $p_a = 0.001$, $p_p = 0.05$, $p_s = 0.005$ and $N = 10^5$, respectively for $p_r = 0$ (top) and $p_r = 0.005$ (bottom). Simulations are averaged over 100 realizations. On the left panels we show the rate of adoption $R_a(t)$ (green), the rate of termination $R_t(t)$ (blue) and the rate of net change $R_n(t) = R_a(t) - R_t(t)$ (red). On the right panels the corresponding cumulative functions are presented. Solid lines denote the associated theoretical curves. Shaded regions depict different regimes bounded by the characteristic times $\tau_0 = 0$, $\tau = \max(R_n(t))$ and $\tau_2 = R_n(t = 0^+)$, as well as $\tau_3 \to \infty$ (top) and $\tau_3=R_n(t=0^-)$ (bottom).}
\label{fig:simSAS_SARS}
\end{figure}

A careful selection of parameter values allows us to simulate various spreading scenarios. Simply by setting $p_r=0$ we can reduce our model into a SIS-like dynamics \cite{Barrat2008} where a non-trivial equilibrium state appears. This is shown in Fig.~S\ref{fig:simSAS_SARS}A, where the system always ends up in a state with equal adoption and termination rates. It is also reflected by the cumulative sum giving the number of active nodes (red curve in Fig.~S\ref{fig:simSAS_SARS}B), which remains constant in the equilibrium state. The number of active users in equilibrium depends on the relative values of the adoption and termination probabilities. The evolution of the adoption process can be separated into three distinct regimes: In the initial regime from time $\tau_0=0$ until $\tau=\max(R_n(t))$ (dark-shaded region), the adoption spreads with an increasing speed and reaches the largest possible number of users. In the second regime from $\tau$ until $\tau_2=R_n(t=0^+)$ (light-shaded regime), the adoption spreading slows down as it becomes controlled by an increasing termination rate. In the third regime from $\tau_2$ until $\tau_3\to\infty$ (white region) the systems stays in equilibrium, which means no change in the account network size since the same number of users adopt and terminate in each time step.

A different scenario takes place if we allow agents to enter a removed state (see Fig.~S\ref{fig:simSAS_SARS}C). In this case the spreading process always reaches a trivial final state where no susceptible nodes remain and no more adoption can happen in the network. Its evolution can also be divided in three typical regimes, of which the first two are similar to the previous scenario. However, as in here termination to a removed state is also possible, the third regime spanning from $\tau_2$ until $\tau_3 = R_n(t = 0^-)$ (white region) starts at a crossover point when termination becomes dominant, the adoption network starts to reduce and approaches the trivial final state. The same scenario can be followed in Fig.~S\ref{fig:simSAS_SARS}D where $\tau$ and $\tau_2$ correspond respectively to the first inflection point and to the maximum of the curve measuring the total number of active users (red line).

\subsection{Non-trivial equilibrium states}
\label{sec:equilStates}

The planar system in Eqs.~(\ref{eq:uncorrODEs1})-(\ref{eq:uncorrODEs2}) has a single fixed point $(0, 0)$ for nonzero $p_r$, meaning that the dynamics always ends up in a state where all individuals are removed and will never use the product again. A different spreading scenario can be obtained by setting $p_r = 0$, which leads to a non-trivial equilibrium state where the rates of adoption and termination are different from zero. Indeed, for $r = 0$ and $a + s = 1$ $\forall\, t$ the 0-clines of Eq.~(\ref{eq:0clines}) become equal,
\begin{equation}
\label{eq:equal0clines}
s_a = s_s = \frac{p_s a}{p_a + p_{pk}(1 - p_a)a},
\end{equation}
and the final state $(a^{\infty}, s^{\infty})$ can be found by inserting Eq.~(\ref{eq:equal0clines}) in the normalization condition $s^{\infty} = 1 - a^{\infty}$. The resulting quadratic equation gives,
\begin{equation}
\label{eq:adoptSols}
a^{\infty}_{\pm} = \frac{1}{2p_{pk}(1 - p_a)} \left[ p_{pk}(1 - pa) - p_a - p_s \pm \sqrt{ [p_{pk}(1 - p_a) - p_a - p_s]^2 + 4p_a p_{pk} (1 - p_a) } \right].
\end{equation}
Since $4p_a p_{pk} (1 - p_a) > 0$ for $p_a, p_{pk} \neq 0$, we have $a^{\infty}_- < 0$ and the stationary probability of adoption can only be $a^{\infty} = a^{\infty}_+ \neq 0$. Moreover, the condition $da/dt = 0$ defining the final state implies that the stationary rates of adoption and termination are equal to an equilibrium rate $R^{\infty} = p_s a^{\infty}$, which after some algebra can be written as,
\begin{equation}
\label{eq:equilRate}
R^{\infty} = \frac{p_s}{2} \left( 1 - \frac{p_a + p_s}{p_{pk}(1 - p_a)} + \sqrt{ \left[ 1 - \frac{p_a + p_s}{p_{pk}(1 - p_a)} \right]^2 + \frac{4 p_a}{p_{pk}(1 - p_a)} } \right).
\end{equation}

The condition $p_r = 0$ simplifies the system~(\ref{eq:uncorrODEs1})-(\ref{eq:uncorrODEs2}) into the single autonomous equation $da/dt = [p_a + p_{pk}(1 - p_a)a](1 - a) - p_s a$. After integrating directly and rearranging terms with the help of Eq.~(\ref{eq:adoptSols}) we obtain,
\begin{equation}
\label{eq:solvedAdopt}
a(t) = a^{\infty}_+ \frac{e^{t/\tau_c} - 1}{e^{t/\tau_c} - a^{\infty}_+/a^{\infty}_-},
\end{equation}
where we have defined a characteristic time $\tau_c = 1/\sqrt{ [p_{pk}(1 - p_a) - p_a - p_s]^2 + 4p_a p_{pk}(1 - p_a) }$. The explicit solution in Eq.~(\ref{eq:solvedAdopt}) can be used to derive analytical expressions for the temporal evolution of the rates $R_a$ and $R_t$ of Eq.~(\ref{eq:ratesDef}), which in turn lead to expressions of important moments in the dynamics like the times $t_a$ and $t_n$, defined respectively as the times when the rates $R_a$ and $R_n = R_a - R_t$ are maximal. Explicitly,
\begin{equation}
\label{eq:explicitTimes}
t_a = \tau_c \ln \left( \frac{a^{\infty}_+}{a^{\infty}_-} \cdot \frac{-p_s - 1/\tau_c}{-p_s + 1/\tau_c} \right)
\qquad \text{and} \qquad
t_n = \tau_c \ln \left( -\frac{p_{pk}(1 - p_a) - p_a - p_s + 1/\tau_c}{p_{pk}(1 - p_a) - p_a - p_s - 1/\tau_c} \right).
\end{equation}

Since $da/dt$ is autonomous, we can also find the maximal rates of adoption and net change ($R^*_a = R_a(t_a)$ and $R^*_n = R_n(t_n)$) through a quicker route. The condition $0 = dR_a/dt |_{t = t_a}$ implies $2a(t_a) = 1 - p_a/[p_{pk}(1 - p_a)]$, so we can substitute in the left side of Eq.~(\ref{eq:ratesDef}) and write,
\begin{equation}
\label{eq:finalRateAdopt}
R^*_a = \frac{p_a}{4} \left( 1 + \frac{p_a}{p_{pk}(1 - p_a)} \right) \left( 1 + \frac{p_{pk}(1 - p_a)}{p_a} \right).
\end{equation}
Similarly, for the net rate we have $2a(t_n) = 1 - (p_a + p_s)/[p_{pk}(1 - p_a)]$ and,
\begin{equation}
\label{eq:finalNetRate}
R^*_n = \frac{1}{4} \left( 1 + \frac{p_a + p_s}{p_{pk}(1 - p_a)} \right) [p_a + p_{pk}(1 - p_a) - p_s] - \frac{p_s}{2} \left( 1 - \frac{p_a + p_s}{p_{pk}(1 - p_a)} \right).
\end{equation}

Eqs.~(\ref{eq:equilRate}), (\ref{eq:finalRateAdopt}) and (\ref{eq:finalNetRate}) define a non-linear algebraic system between the rates $\{ R^{\infty}, R^*_a, R^*_n \}$ and the parameters $\{ p_a, p_{pk}, p_s \}$, one that may be used to solve the inverse problem of determining appropriate parameters in terms of measured rates. The resulting parameters and their errors give in this way estimated areas for the time evolution of the rates in the system.

\section{Empirical fits}
\label{sec:empfits}

To obtain the best model fit of the empirical rates we apply a bounded non-linear least square method and fit the binned $R_a(t)$, $R_t(t)$, and $R_n(t)$ curves simultaneously. As discussed in the main text, the model is determined by the parameter set $\{ p_a, p_p, p_r, \widetilde{p_s}, \widetilde{\langle k \rangle} \}$. There we present fitting results where $\{ \widetilde{p_s}, \widetilde{\langle k \rangle} \}$ are always estimated from the data and $\{ p_a, p_p, p_r \}$ are considered as free parameters. However, in some countries the data allows for the empirical determination of $\widetilde{p_a}$ as well. Out of the $34$ investigated countries, we could estimate this parameter empirically for $11$ countries and perform the fitting with the parameter set $\{ \widetilde{p_a}, p_p, p_r, \widetilde{p_s}, \widetilde{\langle k \rangle} \}$ of two free ($p_p, p_r$) and three fixed ($\widetilde{p_a}, \widetilde{p_s}, \widetilde{\langle k \rangle}$) parameters.

\subsection{Two vs. three free parameter fits}

\begin{figure}
\centering
\includegraphics[width=135mm]{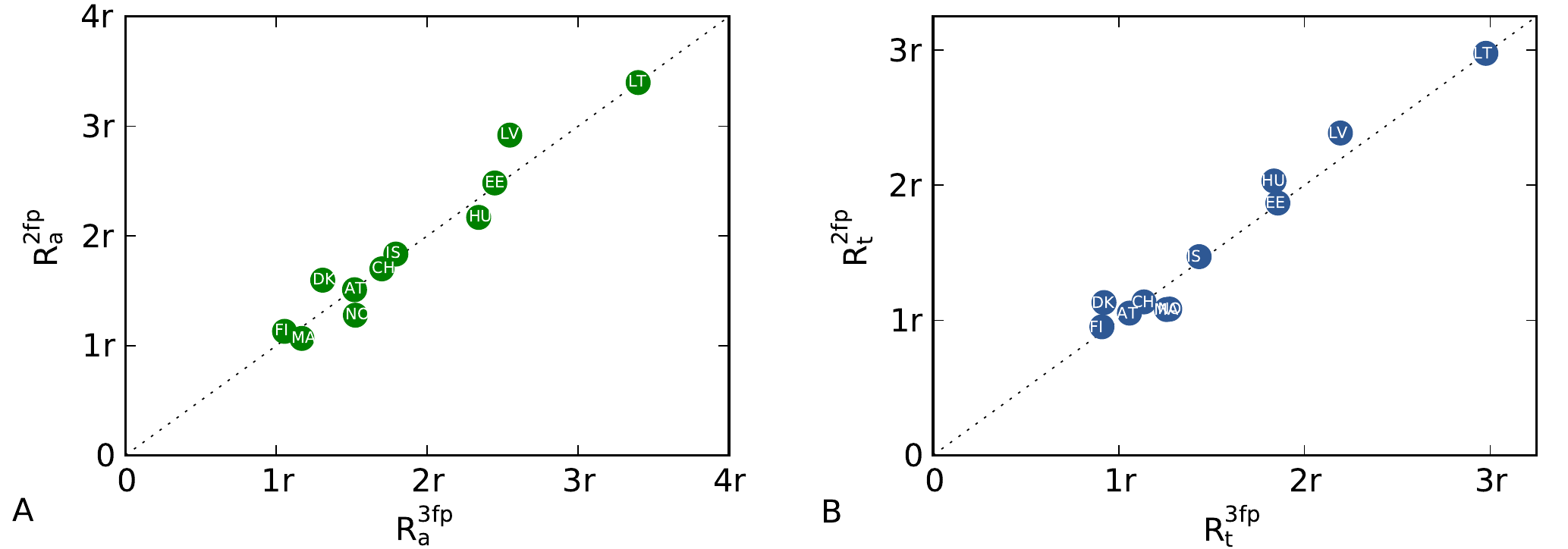}
\caption{Comparison between model predictions. Predictions obtained by fits with three (x-axis) and two (y-axis) free parameters are shown for the 11 different countries where $\widetilde{p_a}$ is empirically determinable. Symbols depict (a) rates of adoption $R_a(t)$ (green) and (b) rates of termination $R_t(t)$ (blue) averaged over the last six months of the observation period, with their corresponding standard deviations as error bars (smaller than symbols). In-symbol letters are country abbreviations, $r$ is an arbitrary linear scaling constant, and the dashed line is a linear function with unit slope.}
\label{fig:23Fit}
\end{figure}

In Fig.~S\ref{fig:23Fit} we compare the predicted modelled rates obtained by fitting the empirical rates with either two or three free parameters. Although some differences appear between the two results, in most countries the corresponding symbols collapse in a line with unit slope and thus assign excellent agreement between the predicted rates. This implies that even fits with three free parameters provide good predictions about the asymptotic evolution of the adoption process.

\subsection{Goodness of fits}

To quantify the quality of the obtained empirical fits we performed the following analysis. The model fit is calculated over the first $5$ years of the data, so that the end of the training period is always fixed. In contrast the beginning of the training period may slightly vary from country to country, depending on the length of the initial transient state. Consequently, a comparable fit quality measure is calculated as follows: The model rates are fitted over the curves $r^e_a$, $r^e_t$ and $r^e_n$, each a group of 10 (or less) binned empirical values of $R_a(t)$, $R_t(t)$ and $R_n(t)$, respectively. We then derive the triplet,
\begin{equation}
r_*=\left(\frac{r_*^e-\overline{r_*^m}}{\text{max}(r_*^e)}\right)^2,
\end{equation}
where $*\in \{a,t,n\}$ and $\overline{r_*^m}$ is the corresponding average model value. Using this triplet we calculate the combined Residual Sum of Squares ($RSS$) per point set as,
\begin{equation}
RSS=\frac{ \sum_{t=n_{min}}^n r_a(t)+r_t(t)+r_n(t)}{n-n_{min}},
\end{equation}
where $n$ is an index of measure points ($\text{max}(n)=10$) and $n_{min}$ is the minimum index of the actual country measure. The obtained $RSS$ values are summarized in Table~S\ref{table:1}.

\begin{table}
\center
\begin{tabular}{|l|l|c|c|}
\hline 
Abbreviation & Country name & RSS (3 free parameters) &  RSS (2 free parameters) \\ 
\hline \hline 
AT & Austria & 0.030547 & 0.030640 \\ \hline 
CH & Switzerland & 0.089469 & 0.089468 \\ \hline 
DK & Denmark & 0.089758 & 0.111208 \\ \hline 
EE & Estonia & 0.083691 & 0.084199 \\ \hline 
FI & Finland & 0.094170 & 0.336367 \\ \hline 
HU & Hungary & 0.075269 & 0.378541 \\ \hline 
IS & Iceland & 0.091920 & 0.045883 \\ \hline 
LV & Latvia & 0.021787  & 0.045755 \\ \hline 
LT & Lithuania & 0.130222 & 0.130222 \\ \hline 
MA & Morocco & 0.178242 & 1.058968 \\ \hline 
NO & Norway & 0.065166 & 0.048559 \\ \hline 
\end{tabular} 
\caption{List of investigated countries with three or two free parameters fits. The first and second columns give the country abbreviation codes and names, while the third and forth column includes the corresponding combined $RSS$ values (defined in section \ref{sec:empfits}) obtained by three and two free parameter fits respectively.}
\label{table:1}
\end{table}

\subsection{Correlations with liberty measures}

\begin{figure}[h!]
\centering
\includegraphics[width=135mm]{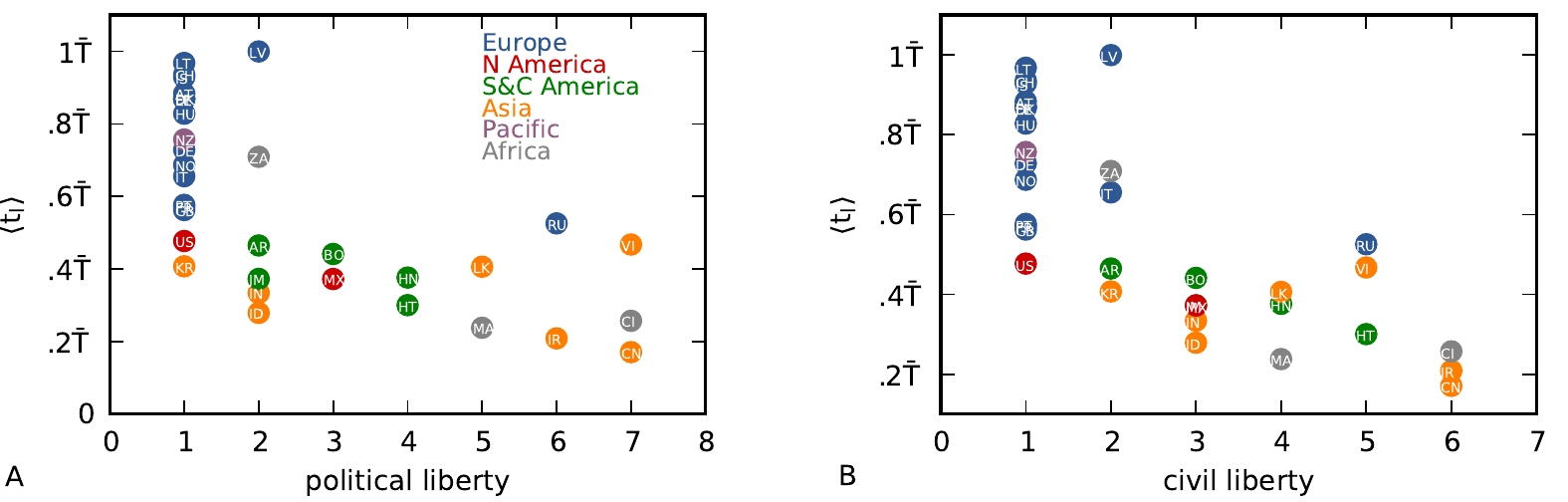}
\caption{Account life time correlations with liberty measures. Average lifetime $\langle t_l \rangle$ of accounts as a function of (a) political and (b) civil liberty measures \cite{Liberty} in various countries (large scores imply weak liberties). $\bar{T}$ is an arbitrary linear scaling constants with time dimension.}
\label{fig:lib}
\end{figure}

As mentioned in the main text, our model of adoption spreading can disclose relevant differences between the adoption dynamics of countries at various levels of societal and economical development. One characteristic indicator in focus is the average lifetime of accounts in a given country defined as $\langle t_l \rangle=\langle t_t-t_a \rangle$, where $t_a$ and $t_t$ are the corresponding registration and termination times. We relate this empirical measure to general liberty measures \cite{Liberty} with results shown in the main text and in Fig.S\ref{fig:lib}. We observe that the weaker the press/political/civil liberty is in a country, the shorter time online accounts are used there. Such observations indicate a quantifiable dependence between the dynamics of innovation spreading and the socio-economic status of a country.

\end{document}